\documentclass[aps,prd,10pt,notitlepage,superscriptaddress,nofootinbib,numbers]{revtex4-2}

\usepackage[utf8]{inputenc}
\usepackage[T1]{fontenc}
\usepackage{lmodern}
\usepackage{microtype}
\usepackage{amsmath,amssymb,amsfonts,mathrsfs}
\usepackage{bm}
\usepackage{graphicx}
\usepackage{float}
\usepackage[dvipsnames]{xcolor}
\usepackage{hyperref}
\hypersetup{colorlinks=true,citecolor=Purple,linkcolor=Purple,urlcolor=Purple,breaklinks=true}
\usepackage[capitalize]{cleveref}
\usepackage{xcolor}

\usepackage{graphicx}
\usepackage{hyperref}
\usepackage{tikz}

\newcommand{\dd}{\mathrm{d}}

\newcommand{\sig}{\sigma}

\newcommand{\Gv}{G_{\rm v}}
\newcommand{\mv}{m_{\rm v}}
\newcommand{\rhop}{\rho_{\rm v}}
\newcommand{\Ueff}{\Upsilon_{\rm eff}}

\definecolor{lime}{HTML}{A6CE39}
\DeclareRobustCommand{\orcidicon}{%
	\begin{tikzpicture}
		\draw[lime, fill=lime] (0,0) 
		circle [radius=0.16] 
		node[white] {{\fontfamily{qag}\selectfont \tiny ID}};
		\draw[white, fill=white] (-0.0625,0.095) 
		circle [radius=0.007];
	\end{tikzpicture}
	\hspace{-2mm}
}

\foreach \x in {A, ..., Z}{%
	\expandafter\xdef\csname orcid\x\endcsname{\noexpand\href{https://orcid.org/\csname orcidauthor\x\endcsname}{\noexpand\orcidicon}}
}

\newcommand\orcidFrancisco{{\href{https://orcid.org/0000-0002-9388-8373}{\orcidicon}}}
\newcommand\orcidJonathan{{\href{https://orcid.org/0000-0001-9291-0893}{\orcidicon}}}
\newcommand\orcidEdson{{\href{https://orcid.org/0000-0001-9929-5977}{\orcidicon}}}

\begin{document}

\title{Thin-shell wormholes in cosmic voids}

\author{Jonathan A. Rebou\c{c}as\orcidJonathan\!\!}
\email{jalvesreboucas@ifce.edu.br}
\affiliation{Instituto Federal de Educa\c{c}\~ao, Ci\^encia e Tecnologia do Cear\'a (IFCE), Iguatu-CE, Brazil}

\author{Edson Otoniel\orcidEdson\!\!}
\email{edson.otoniel@ufca.edu.br}
\affiliation{Universidade Federal do Cariri (UFCA), Instituto de Forma\c{c}\~ao de Educadores - IFE,  R. Oleg\'ario Emidio de Araujo S/N, Brejo Santo - CE, 63.260-000 - Brazil}

%%%%%%%%%%%%%%%%%%%%%%%%%%%%%%%%%%%%%%%%%%%%%%%%%%%%%%%%%%%%%%%% 
\author{Francisco S. N. Lobo\orcidFrancisco\!\!} 
\email{fslobo@ciencias.ulisboa.pt}
\affiliation{Instituto de Astrof\'{i}sica e Ci\^{e}ncias do Espa\c{c}o, Faculdade de Ci\^{e}ncias da Universidade de Lisboa, Edifício C8, Campo Grande, P-1749-016 Lisbon, Portugal}
\affiliation{Departamento de F\'{i}sica, Faculdade de Ci\^{e}ncias da Universidade de Lisboa, Edif\'{i}cio C8, Campo Grande, P-1749-016 Lisbon, Portugal}
%%%%%%%%%%%%%%%%%%%%%%%%%%%%%%%%%%%%%%%%%%%%%%%%%%%%%%%%%%%%%%%%

\date{\today}

\begin{abstract}
	Cosmic voids are underdense regions that can provide an effective large-scale environment with a de Sitter-like gravitational behavior. Motivated by recent black-hole solutions embedded in void density profiles, we construct a symmetric thin-shell wormhole by gluing two copies of the positive-lapse region of a black hole inside a cosmic void. The surface stresses are obtained from the Darmois--Israel junction conditions, and the corresponding null, weak, dominant, and strong energy-condition combinations are written directly in terms of the void mass function and density profile. We further develop the thermodynamics of the static shell, deriving a first law that relates the shell entropy to the black-hole and cosmological-like horizon entropies. We then formulate the radial dynamics of the throat through an effective potential and study the local stability of static configurations when the exotic matter on the shell obeys either a generalized cosmic Chaplygin gas or a modified cosmic Chaplygin gas equation of state. In both models the Chaplygin parameter $B$ is fixed by the static junction condition, so that the remaining stability test is governed by the void geometry and by the equation-of-state parameters. Numerical results reveal that GCCG-supported configurations are generically unstable, whereas MCCG-supported wormholes can be stable for a sufficiently large linear term in the equation of state. The resulting framework connects the de Sitter-like structure of cosmic voids with the standard thin-shell wormhole formalism and provides a starting point for identifying stable or unstable wormhole configurations located between the black-hole and cosmological-like horizons of the void spacetime.
\end{abstract}

\maketitle

\tableofcontents

\section{Introduction}\label{sec:introduction}

A recurrent problem in relativistic astrophysics is how local strong-gravity configurations are affected by the large-scale environment in which they are embedded. Compact objects are usually modeled as isolated systems, but the Universe is structured on scales much larger than galaxies, and this hierarchy can influence the interpretation of horizons, effective potentials, and quasi-local gravitational behavior. Cosmic voids provide a particularly relevant setting for this question. They are extended underdense regions of the cosmic web, surrounded by walls and filaments, and they occupy a substantial fraction of the cosmic volume. Their formation, evolution, and statistical properties have been studied since the first large-scale descriptions of the void network and remain an important probe of structure formation, dark energy, and possible departures from standard gravity~\cite{Zeldovich1982,ElAd1997,Gottlober2003,Colberg2005,deWeygaert2011,Pisani2019,Schuster2023}.

The physical importance of voids is not limited to their low density. Because they correspond to weak-field and low-screening environments, voids can enhance signatures that are difficult to isolate in overdense regions. Their density contrasts, compensation walls, expansion rates, and clustering properties are therefore useful for testing cosmological dynamics and modified-gravity scenarios~\cite{Bos2012,Chan2014,Moretti2026,Tavasoli2026}. A central ingredient in this program is the characterization of the void density profile. The universal profile proposed by Hamaus, Sutter, and Wandelt offers a phenomenological description of the underdense core and surrounding compensation wall, and has become a useful tool for connecting void observations with theoretical modeling~\cite{Hamaus2014}. This profile also permits one to ask whether the matter distribution of a void can be used directly as a source for local relativistic geometries.

This question has recently led to the construction of black-hole solutions inside cosmic voids~\cite{Lustosa2025}. In this approach, a central mass is embedded in the matter distribution associated with the universal void profile, producing a static and spherically symmetric lapse function with two characteristic roots in the relevant parameter range: an inner black-hole horizon and an outer cosmological-like horizon generated by the de Sitter-like behavior of the void environment. The resulting spacetime is not simply an isolated Schwarzschild geometry with a small correction. Instead, the void contribution creates a finite positive-lapse region bounded by horizon-like surfaces, in close analogy with the role played by a positive cosmological constant in Schwarzschild--de Sitter geometries~\cite{LoboCrawford2004,SharifMumtaz2014}. This feature makes the black-hole-in-void metric a natural seed for exploring compact objects whose admissible domain is controlled simultaneously by a local horizon and by a large-scale environmental boundary.

Wormholes provide a natural arena in which this interplay between local geometry and global environment can be investigated. The Einstein--Rosen bridge introduced an early nontraversable connection between distinct regions of spacetime~\cite{EinsteinRosen1935}, while the Morris--Thorne framework established the modern concept of traversable wormholes and emphasized the connection between throat geometry and violations of the classical energy conditions~\cite{MorrisThorne1988}. In general relativity, these violations are closely associated with the flare-out condition and have motivated several strategies for reducing or localizing the required exotic matter~\cite{Hochberg1998,Visser2003,Konoplya2022}. Related approaches have investigated wormholes supported by phantom energy, Casimir sources, dark matter halos, loop-quantum-gravity corrections, and effective contributions from modified gravity~\cite{Sushkov2005,Lobo:2005us,Lobo2009,Lobo2012,Boehmer2012,Harko2013,Kanti2012,Garattini2019,Rahaman2016,Cruz2024,Santos2024,Radhakrishnan2024}. Thin-shell wormholes offer one of the most economical realizations of this idea: two copies of a seed spacetime are glued at a timelike hypersurface, so that the nontrivial matter source is concentrated on the shell rather than distributed through the bulk~\cite{Visser1989,VisserBook,Lobo:2004id,Lobo:2004rp,Lobo:2005zu}.

The dynamics of thin-shell wormholes is determined by the geometry of the seed spacetime and by the matter equation of state on the shell. The Darmois--Israel--Lanczos junction formalism relates the jump of the extrinsic curvature to the surface stress tensor and provides the surface energy density and tangential pressure supporting the throat~\cite{Darmois1927,Lanczos1924,Israel1966}. Linearized stability around static configurations was first developed for Schwarzschild thin-shell wormholes by Poisson and Visser~\cite{PoissonVisser1995}, and was later extended to charged, cosmological, higher-dimensional, modified-gravity, and quantum-corrected backgrounds~\cite{Eiroa2004ChargedStability,Eiroa2008SphericalStability,Eiroa2009,Eiroa2012,Varela2015,Tangphati2020,Tsukamoto2021,Javed2023,Eiroa2024,Eiroa2026,Javed2024}. Further developments have considered regular black-hole seeds, rotating shells, higher-curvature corrections, lower- and higher-dimensional settings, and thermodynamic aspects of thin-shell wormholes~\cite{Thibeault2006,Richarte2007,Dias2010,Mazharimousavi2010,Garcia2012,Halilsoy2014,Mazharimousavi2014,Eiroa2016,Tsukamoto2018,Forghani2019,Kokubu2020,Lobo2020,Bejarano2021}. In parallel, Chaplygin-type fluids have often been used as phenomenological models for the exotic matter at the throat~\cite{Lobo:2005vc}, allowing one to close the dynamical system and test the sign of the second derivative of the effective potential~\cite{GonzalezDiaz2003,SadeghiFarahani2013,SharifMumtaz2014}.
More recently, a unified thermodynamic framework for thin-shell wormholes has been developed~\cite{Lobo:2026vrn}, establishing a generalised first law and second law, and relating the shell's Unruh temperature to Hawking-like particle creation~\cite{Lobo:2026qhw}. These advances motivate the thermodynamic analysis of the void-embedded wormhole presented below.

The possibility of connecting wormhole physics with cosmic void environments has also begun to appear in recent studies. Wormhole geometries supported by void matter profiles and optical signatures of wormholes in void backgrounds have been investigated in different gravitational settings~\cite{Errehymy2025,Errehymy2026}. These works indicate that voids can affect both the geometry and phenomenology of wormhole spacetimes. However, a different and complementary question remains open: whether the black-hole-in-void geometry itself can be used as a seed spacetime for a thin-shell wormhole, and how the two-horizon structure of the void spacetime constrains the shell stresses, energy conditions, and stability domain. This question is distinct from constructing a continuous Morris--Thorne wormhole supported by a bulk fluid, because the exotic matter is confined to a junction surface and the admissible throat radius must lie inside the positive-lapse interval of the seed black-hole geometry.

In this work, we construct a symmetric thin-shell wormhole inside a cosmic void by applying the cut-and-paste procedure to the black-hole-in-void metric generated by the universal void density profile, studied in Ref. \cite{Lustosa2025}. The throat is placed in the region between the inner black-hole horizon and the outer cosmological-like horizon, where the lapse function is positive. We derive the induced metric, the unit normal, and the extrinsic-curvature components, and then use the Darmois--Israel formalism to obtain the surface energy density and tangential pressure. The null, weak, dominant, and strong energy-condition combinations are written explicitly in terms of the void mass function and density profile. We then develop a thermodynamic description of the static shell, defining the relevant temperatures, deriving a first law, and establishing a direct relation between the shell entropy and the horizon entropies. We then formulate the radial dynamics of the shell through an effective potential and analyze the local stability of static configurations supported by generalized cosmic Chaplygin gas (GCCG) and modified cosmic Chaplygin gas (MCCG) equations of state.

The main motivation is to identify how the large-scale underdense environment modifies the standard thin-shell picture. In asymptotically flat thin-shell wormholes, the throat is usually constrained only by the black-hole horizon or by the regularity of the seed geometry. In the present case, the void profile introduces an additional outer boundary and a de Sitter-like contribution that can strongly affect both the available static interval and the pressure regime on the shell. Thus, the construction provides a controlled way to connect three ingredients that are usually treated separately: cosmic-void density profiles, black-hole horizon structure, and thin-shell wormhole stability.

This paper is organized as follows. In Sec.~\ref{sec:void_geometry}, we review the void density profile, the associated mass function, and the black-hole-in-void lapse function. In Sec.~\ref{sec:thin_shell}, we construct the thin-shell wormhole and present the induced geometry, normal vector, and extrinsic curvature. Section~\ref{sec:surface_energy} derives the surface stress tensor and discusses the energy-condition combinations on the shell. In Sec.~\ref{sec:thermo}, we examine the thermodynamics of the thin‑shell wormhole, including the first law and the connection to the horizon entropies. Section~\ref{sec:dynamics} then reduces the radial motion to a one-dimensional effective potential and analyzes the linear stability for the Chaplygin-type equations of state. Finally, Sec.~\ref{sec:conclusion} summarizes our results and discusses possible extensions. We use geometrized units $G=c=\hbar=k_B=1$.

\section{Black holes inside cosmic voids}\label{sec:void_geometry}

\subsection{Void density profile and mass function}

We start from the static and spherically symmetric line element
\begin{equation}
	\dd s^2=-\Gv(r)\dd t^2+\frac{\dd r^2}{\Gv(r)}+r^2\left(\dd\theta^2+\sin^2\theta\dd\phi^2\right),
	\label{eq:void_metric}
\end{equation}
where $\Gv(r)$ is the lapse function generated by a central mass embedded in a cosmic-void density profile. The void matter distribution is modeled through the universal density profile of Hamaus, Sutter, and Wandelt~\cite{Hamaus2014}, as used in Ref.~\cite{Lustosa2025}:
\begin{equation}
	\rhop(r)=\rho_0\left[1+\delta_c\frac{1-\left(r/r_s\right)^\alpha}{1+\left(r/r_v\right)^\beta}\right].
	\label{eq:void_density}
\end{equation}
Here $\rho_0$ is the mean background density, $\delta_c<0$ measures the depth of the void, $r_s$ is the scale at which the density reaches the background value, and $r_v$ is the void radius. The parameters $\alpha$ and $\beta$ control the inner slope and the compensation wall. Following the phenomenological calibration adopted in Ref.~\cite{Lustosa2025} from the universal void profile of Hamaus, Sutter, and Wandelt~\cite{Hamaus2014}, the shape parameters can be written as functions of the ratio $r_s/r_v$:
\begin{equation}
	\alpha(r_s)=-2\left(\frac{r_s}{r_v}-2\right),
	\label{eq:alpha_void}
\end{equation}
\begin{equation}
	\beta(r_s)=
	\begin{cases}
		17.5\,\dfrac{r_s}{r_v}-6.5, & r_s/r_v<0.91,\\[6pt]
		-9.8\,\dfrac{r_s}{r_v}+18.4, & r_s/r_v>0.91.
	\end{cases}
	\label{eq:beta_void}
\end{equation}

\begin{figure}[htp!]
	\centering
	\includegraphics[width=0.65\linewidth]{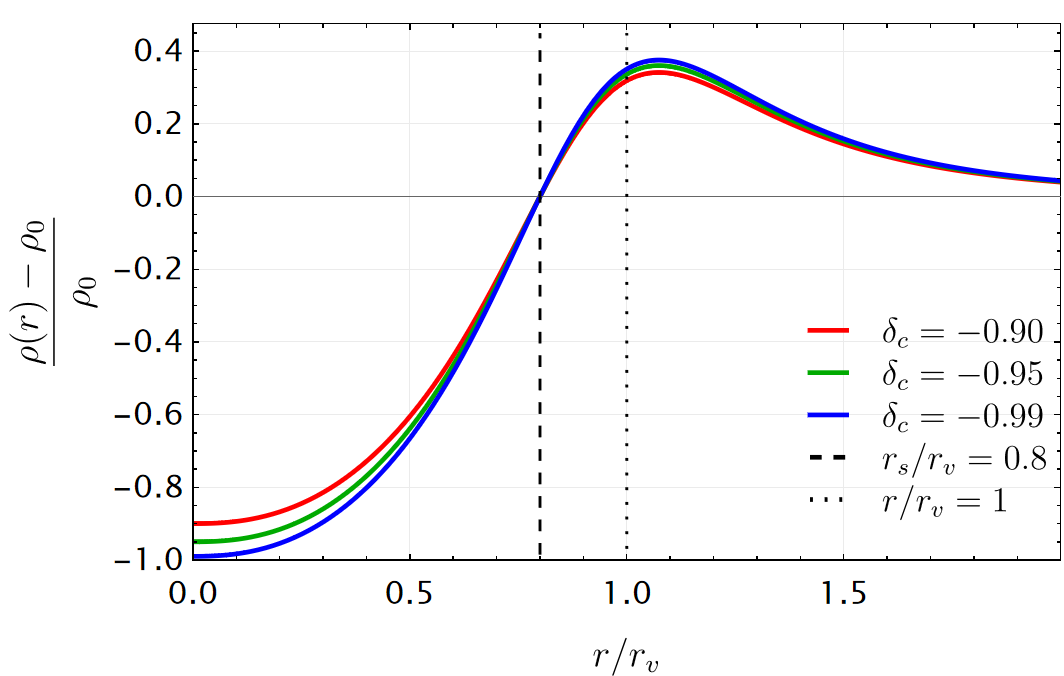}
	\caption{Normalized density contrast $(\rho(r)-\rho_0)/\rho_0$ as a function of the dimensionless radius $r/r_v$ for the universal cosmic-void profile. The curves correspond to $\delta_c=-0.90$, $-0.95$, and $-0.99$, with $r_s=80$, $r_v=100$, $\alpha=2.4$, and $\beta=7.5$. The dashed and dotted vertical lines indicate $r_s/r_v=0.8$ and $r/r_v=1$, respectively.}
	\label{fig:void_density_contrast}
\end{figure}

Figure~\ref{fig:void_density_contrast} shows the normalized density contrast $(\rho(r)-\rho_0)/\rho_0$ associated with the universal cosmic-void profile. Near the center, this quantity approaches $\delta_c$, so that larger values of $|\delta_c|$ correspond to deeper underdense cores. All curves cross zero at $r=r_s$, marked by the dashed vertical line, indicating the radius at which the local density becomes equal to the background density $\rho_0$. For $r_s<r\lesssim r_v$, the normalized contrast becomes positive, signaling the presence of the compensation wall surrounding the void. The dotted vertical line marks the void radius $r=r_v$, around which the overdense shell is still visible. At larger radii, the curves decrease toward zero, showing the gradual recovery of the average cosmic density outside the void. Therefore, the figure highlights the three characteristic regions of the profile: the underdense core, the transition at $r_s$, and the compensated wall near the void boundary.

For the metric in Eq.~\eqref{eq:void_metric}, the function $\mv(r)$ is obtained from the $tt$ component of Einstein's equations after introducing the Schwarzschild-like parametrization used in Eq.~\eqref{eq:void_lapse}. In this form, one has
\begin{equation}
	\mv'(r)=4\pi r^2\rhop(r).
	\label{eq:void_mass_differential}
\end{equation}
The integration constant $M$ is interpreted as the ADM mass of the central object, so that the full mass function reads
\begin{equation}
	\mv(r)=M+4\pi\int_0^r u^2\rhop(u)\,\dd u .
	\label{eq:void_mass_integral}
\end{equation}
For the density profile in Eq.~\eqref{eq:void_density}, this gives
\begin{equation}
	\begin{aligned}
		\mv(r)=M+\frac{4\pi}{3}r^3\rho_0\Bigg[1
		&+\delta_c\,{}_2F_1\left(1,\frac{3}{\beta};\frac{3+\beta}{\beta};-\left(\frac{r}{r_v}\right)^\beta\right)\\
		&-\frac{3\delta_c}{3+\alpha}\left(\frac{r}{r_s}\right)^\alpha
		{}_2F_1\left(1,\frac{3+\alpha}{\beta};\frac{3+\alpha+\beta}{\beta};-\left(\frac{r}{r_v}\right)^\beta\right)\Bigg],
	\end{aligned}
	\label{eq:void_mass_function}
\end{equation}
where ${}_2F_1(a,b;c;z)$ denotes the Gaussian hypergeometric function.
The lapse function is written as
\begin{equation}
	\Gv(r)=1-\frac{2\mv(r)}{r}.
	\label{eq:void_lapse}
\end{equation}

\subsection{Horizons and de Sitter-like limit}

The horizons are determined by the positive roots of
\begin{equation}
	\Gv(r)=0.
	\label{eq:void_horizon_condition}
\end{equation}
In the parameter range of interest, the geometry has an inner black-hole horizon $r_h$ and an outer cosmological-like horizon $r_{++}$, \cite{Lustosa2025}. The static region relevant for the thin-shell construction is therefore
\begin{equation}
	r_h<r<r_{++},
	\qquad
	\Gv(r)>0.
	\label{eq:static_void_domain}
\end{equation}
If the wormhole is required to be located inside the void, one additionally imposes $r<r_v$, and in numerical applications it is useful to restrict the throat to the interval $r_h<\chi_0<\min(r_{++},r_v)$.

The de Sitter-like character of the void metric follows from the intermediate regime far from the central singularity and well inside the void wall, where the density is nearly constant. For $0\ll r\ll r_s<r_v$, the density approaches $\rho_0(1+\delta_c)$ and the lapse becomes
\begin{equation}
	\Gv(r)\simeq 1-\frac{2M}{r}-\frac{8\pi}{3}\rho_0(1+\delta_c)r^2
	=1-\frac{2M}{r}-\frac{\Lambda_{\rm v}}{3}r^2,
	\label{eq:desitter_like_with_mass}
\end{equation}
where
\begin{equation}
	\Lambda_{\rm v}=8\pi\rho_0(1+\delta_c).
	\label{eq:lambda_void}
\end{equation}
When the central term is negligible in comparison with the void term, Eq.~\eqref{eq:desitter_like_with_mass} reduces to the de Sitter-like form
\begin{equation}
	\Gv(r)\simeq1-\frac{\Lambda_{\rm v}}{3}r^2.
	\label{eq:desitter_like_void}
\end{equation}
Thus, the outer horizon of the static patch is the analogue of a cosmological horizon induced by the effective low-density environment of the void.

\subsection{Lapse function profiles}

\begin{figure}[htp!]
	\centering
	\includegraphics[width=0.49\linewidth]{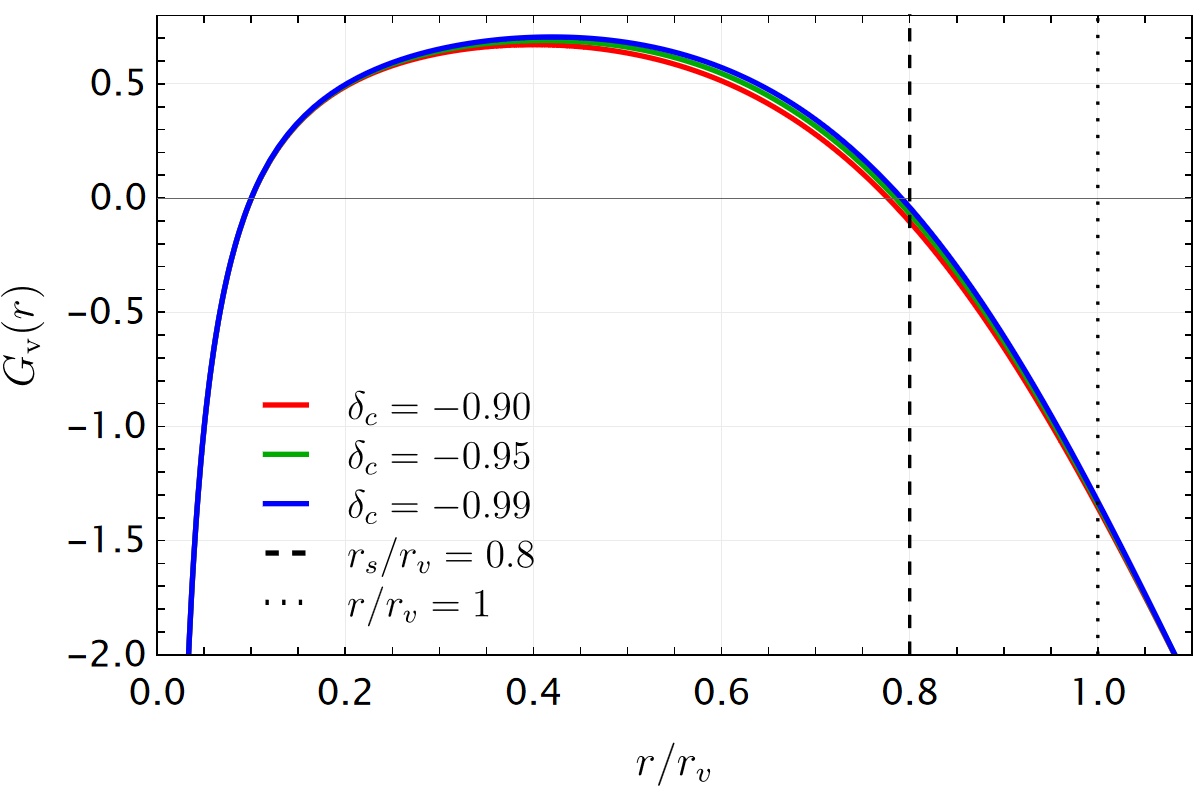}
	\includegraphics[width=0.49\linewidth]{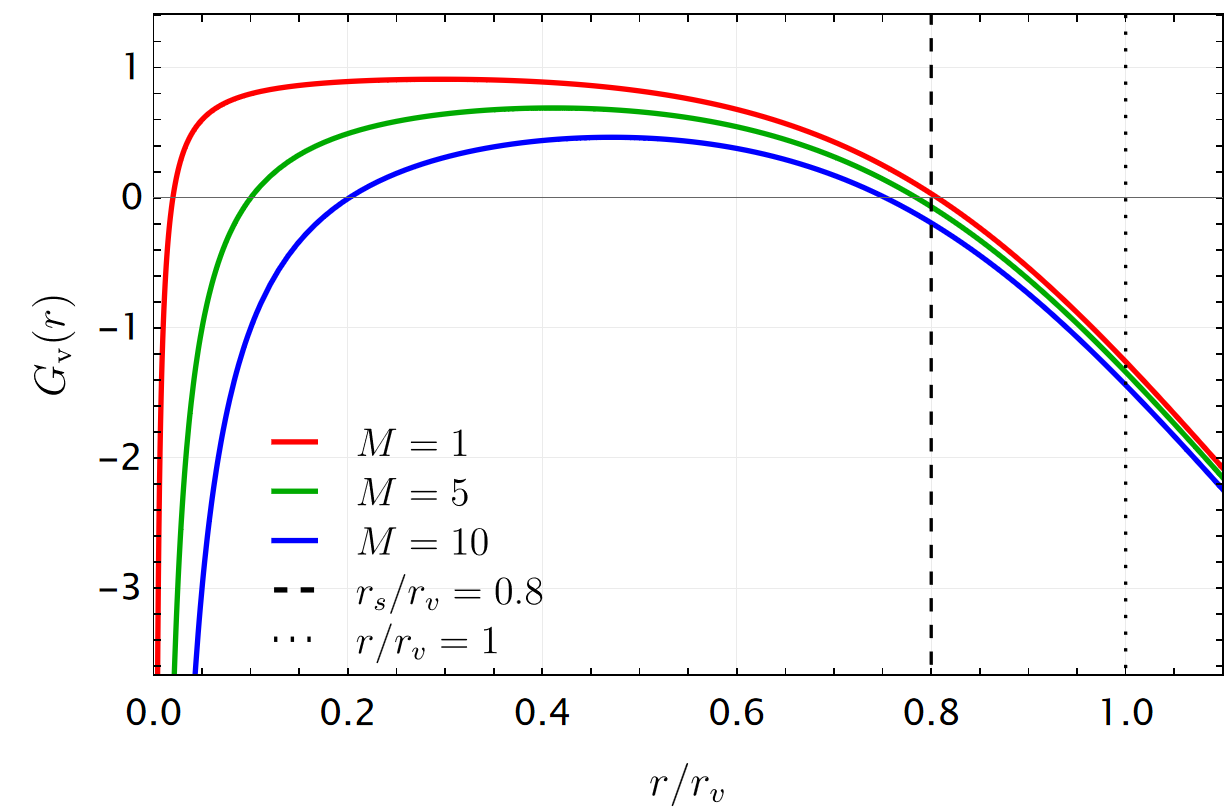}
	\caption{Void lapse function $\Gv(r)$ as a function of the dimensionless radius $r/r_v$. In the left panel, the mass is fixed at $M=5$ and the curves correspond to $\delta_c=-0.90$, $-0.95$, and $-0.99$. In the right panel, the density contrast is fixed at $\delta_c=-0.95$ and the curves correspond to $M=1$, $5$, and $10$. In both panels, we use $r_s=80$, $r_v=100$, $\rho_0=3\times10^{-5}$, $\alpha=2.4$, and $\beta=7.5$. The dashed and dotted vertical lines indicate $r_s/r_v=0.8$ and $r/r_v=1$, respectively.}
	\label{fig:void_lapse}
\end{figure}

Figure~\ref{fig:void_lapse} displays the behavior of the void lapse function $\Gv(r)$ in terms of the dimensionless radius $r/r_v$. In both panels, the positive-lapse domain is bounded by two roots of $\Gv(r)$: an inner black-hole horizon and an outer cosmological-like horizon. This is the same qualitative structure found in Ref.~\cite{Lustosa2025}, where the metric coefficient develops two horizons, one associated with the central black hole and the other with the de Sitter-like behavior induced by the void environment. In the left panel, where $M=5$ is fixed, changing $\delta_c$ produces only a small displacement of the inner horizon. This occurs because the near-central region is dominated by the Schwarzschild-like contribution $-2M/r$, so the details of the void density profile act only as a small correction near $r_h$. The effect of $\delta_c$ is more visible near the outer root: as the void becomes emptier, with $\delta_c\to -1$, the cosmological-like horizon moves outward, in agreement with the trend discussed in Ref.~\cite{Lustosa2025}.

The right panel complements this behavior by fixing $\delta_c=-0.95$ and varying the central mass. Increasing $M$ lowers the lapse function over the whole interval, shifts the inner horizon to larger radii, and brings the outer cosmological-like horizon inward. Thus, the available static region between the two roots becomes narrower as the central mass increases. This trend is consistent with the horizon analysis of Ref.~\cite{Lustosa2025}, where the black-hole and outer horizons approach each other as the mass approaches a critical value, eventually merging in a Nariai-like configuration. For the thin-shell construction, this behavior is important because the throat must be placed inside the positive-lapse interval. Therefore, the admissible static radius is not merely constrained by being outside the black-hole horizon, but by the finite domain $r_h<\chi_0<r_{++}$. In the parameter range shown, the outer horizon lies close to, or slightly below, the scale $r_s/r_v=0.8$, indicating that the thin shell is naturally restricted to the inner underdense region of the void before the compensation wall becomes dominant.

\section{Thin-shell wormhole construction}\label{sec:thin_shell}

\subsection{Cut-and-paste and induced geometry}

We construct the thin-shell wormhole by applying the cut-and-paste procedure to the static region of the void geometry. The bulk spacetime is described by the coordinates $x^\mu_\pm=(t_\pm,r_\pm,\theta,\phi)$ on two identical copies $\mathcal{M}^+$ and $\mathcal{M}^-$ of the seed manifold. Since the void geometry possesses an inner black-hole horizon and an outer de Sitter-like horizon, the construction must be performed inside the positive-lapse domain. Therefore, the shell trajectory must satisfy
\begin{equation}
	r_h<\chi(\tau)<r_{++},
	\qquad
	\Gv[\chi(\tau)]>0.
	\label{eq:shell_positive_lapse_domain}
\end{equation}
If the wormhole throat is required to remain inside the void, the additional restriction $\chi(\tau)<r_v$ is imposed. In the static case, this reduces to $r_h<\chi_0<\min(r_{++},r_v)$.

Following the standard thin-shell construction, the regions
\begin{equation}
	\mathcal{M}^{\pm}=\left\{x^\mu_\pm\,\big|\, r_h<\chi(\tau)\leq r_\pm<r_{++}\right\}
	\label{eq:bulk_regions}
\end{equation}
are taken from the two identical copies of the void spacetime and are joined at their common timelike boundary. The two sides are matched with $\mathbb Z_2$ symmetry, which is the origin of the simple relation between the extrinsic curvatures and of the negative sign of the surface energy density obtained in the next section. The resulting geodesically complete manifold is
\begin{equation}
	\mathcal{M}=\mathcal{M}^+\cup\mathcal{M}^-,
	\label{eq:joined_manifold}
\end{equation}
with the junction hypersurface identified as the wormhole throat. The shell is described by
\begin{equation}
	\Sigma:\quad R(r,\tau)=r-\chi(\tau)=0,
	\label{eq:shell_hypersurface}
\end{equation}
where $\chi(\tau)$ is the radial position of the throat and $\tau$ is the proper time measured by an observer comoving with the shell. Thus, the function $\chi(\tau)$ is the dynamical degree of freedom of the throat, and its radial evolution will later be determined from the junction conditions and from the matter equation of state on the shell.

The intrinsic coordinates on the hypersurface are
\begin{equation}
	\xi^i=(\tau,\theta,\phi),
	\label{eq:intrinsic_coordinates}
\end{equation}
while the embedding of the shell in each bulk copy is given by
\begin{equation}
	x^\mu_\pm(\tau,\theta,\phi)=\left(t_\pm(\tau),\chi(\tau),\theta,\phi\right).
	\label{eq:shell_embedding}
\end{equation}
The induced metric, or first fundamental form, is obtained by projecting the bulk metric onto the hypersurface,
\begin{equation}
	h_{ij}^{\pm}=
	g^{\pm}_{\mu\nu}
	\frac{\partial x^\mu_\pm}{\partial \xi^i}
	\frac{\partial x^\nu_\pm}{\partial \xi^j}.
	\label{eq:induced_metric_definition}
\end{equation}
The Darmois matching condition~\cite{Darmois1927,VisserBook} requires the induced geometry to be continuous across the shell, $h_{ij}^+=h_{ij}^-\equiv h_{ij}$, which is the necessary and sufficient condition for the absence of a distributional Riemann tensor at the junction. With the proper-time parametrization on $\Sigma$, the induced line element becomes
\begin{equation}
	\dd s^2_\Sigma
	=
	h_{ij}\dd \xi^i\dd \xi^j
	=
	-\dd\tau^2+\chi^2(\tau)\left(\dd\theta^2+\sin^2\theta\dd\phi^2\right).
	\label{eq:induced_metric}
\end{equation}
For a static throat, $\chi(\tau)=\chi_0$, the shell is located at a fixed radius inside the static void patch, whereas for a dynamical throat the condition in Eq.~\eqref{eq:shell_positive_lapse_domain} must hold along the whole shell trajectory.

\subsection{Normal vector and extrinsic curvature}

The proper-time normalization of the shell four-velocity gives
\begin{equation}
	-\Gv(\chi)\dot t_\pm^2+\frac{\dot\chi^2}{\Gv(\chi)}=-1,
	\qquad
	\dot t_\pm=\frac{\sqrt{\Gv(\chi)+\dot\chi^2}}{\Gv(\chi)}.
	\label{eq:proper_time_normalization}
\end{equation}
In a neighborhood of the timelike hypersurface, one may introduce Gaussian normal coordinates $(\ell,\xi^i)$, where $\ell$ is the proper distance measured along geodesics orthogonal to $\Sigma$ and the shell is located at $\ell=0$. In these coordinates, the normal direction is separated from the intrinsic coordinates on the shell, and the unit normal covector is locally proportional to the gradient of the hypersurface function. Therefore, for the surface \eqref{eq:shell_hypersurface}, the outward-oriented unit normal covector can be written as the normalized gradient of $R$,
\begin{equation}
	\eta_\mu^\pm
	=
	\pm\left|g^{\alpha\beta}\frac{\partial R}{\partial x^\alpha}
	\frac{\partial R}{\partial x^\beta}\right|^{-1/2}
	\frac{\partial R}{\partial x^\mu}
	=
	\pm\left(-\dot\chi,\frac{\sqrt{\Gv(\chi)+\dot\chi^2}}{\Gv(\chi)},0,0\right).
	\label{eq:normal_covector}
\end{equation}
The sign distinguishes the outward orientation on each copy of the bulk. The normal satisfies $\eta_\mu u^\mu=0$ and $\eta_\mu\eta^\mu=1$, where $u^\mu$ is the unit timelike four-velocity of an observer comoving with the shell. Thus, the first condition expresses the orthogonality between the shell trajectory and the normal direction.

The extrinsic curvature can be written as the projection of the covariant derivative of the unit normal onto the hypersurface,
\begin{equation}
	K_{ij}^{\pm}
	=
	e^\alpha_i e^\beta_j \nabla_\alpha \eta_\beta^\pm,
	\qquad
	e^\mu_i=\frac{\partial x^\mu}{\partial \xi^i},
	\label{eq:extrinsic_covariant_definition}
\end{equation}
where $e^\mu_i$ are the tangent basis vectors to $\Sigma$, with $e^\mu_\tau=u^\mu$ along the shell trajectory, and $\eta_\mu^\pm$ is the unit normal covector defined in Eq.~\eqref{eq:normal_covector}. Equivalently, using the embedding functions $x^\mu(\xi^i)$, this definition becomes
\begin{equation}
	K_{ij}^{\pm}=-\eta_\mu^{\pm}\left(\frac{\partial^2x^\mu}{\partial\xi^i\partial\xi^j}
	+\Gamma^\mu_{\alpha\beta}\frac{\partial x^\alpha}{\partial\xi^i}
	\frac{\partial x^\beta}{\partial\xi^j}\right).
	\label{eq:extrinsic_definition}
\end{equation}
Geometrically, $K_{ij}^{\pm}$ measures how the unit normal to the shell changes when it is transported along directions tangent to the hypersurface. It therefore encodes how the shell is embedded in each copy of the bulk spacetime.
In an orthonormal basis on the shell, defined by $\partial_{\hat\tau}=u^\mu\partial_\mu$, $\partial_{\hat\theta}=\chi^{-1}\partial_\theta$, $\partial_{\hat\phi}=(\chi\sin\theta)^{-1}\partial_\phi$, the independent nonzero components are
\begin{equation}
	K_{\hat\tau\hat\tau}^{\pm}=\mp\frac{\Gv'(\chi)+2\ddot\chi}{2\sqrt{\Gv(\chi)+\dot\chi^2}},
	\label{eq:K_tautau}
\end{equation}
\begin{equation}
	K_{\hat\theta\hat\theta}^{\pm}=K_{\hat\phi\hat\phi}^{\pm}=\pm\frac{\sqrt{\Gv(\chi)+\dot\chi^2}}{\chi}.
	\label{eq:K_angular}
\end{equation}
For $\chi=\chi_0$, $\dot\chi=\ddot\chi=0$, these reduce to
\begin{equation}
	K_{\hat\tau\hat\tau}^{\pm}\big|_0=\mp\frac{G_{{\rm v}0}'}{2\sqrt{G_{{\rm v}0}}},
	\qquad
	K_{\hat\theta\hat\theta}^{\pm}\big|_0=K_{\hat\phi\hat\phi}^{\pm}\big|_0=\pm\frac{\sqrt{G_{{\rm v}0}}}{\chi_0},
	\label{eq:K_static}
\end{equation}
where $G_{{\rm v}0}=\Gv(\chi_0)$ and primes are evaluated at $r=\chi_0$.

\section{Surface tensor and energy conditions}\label{sec:surface_energy}

\subsection{Surface stresses from the Lanczos equation}

The construction described in Sec.~\ref{sec:thin_shell} produces a timelike hypersurface $\Sigma$ that acts as the wormhole throat. The bulk geometry is regular on each side of the shell, but the gluing procedure may introduce a non-smooth behavior in the normal derivatives of the metric at $\Sigma$. The matter needed to support the wormhole is therefore not distributed throughout the bulk, but is concentrated on the shell as an intrinsic surface layer. Since the shell is a $(2+1)$-dimensional hypersurface, its stress tensor has only components tangent to $\Sigma$.

We write the surface stress--energy tensor in an orthonormal basis adapted to the shell. Hatted indices refer to the local frame carried by observers comoving with the throat, so that the induced metric on the shell is locally Minkowskian, $h_{\hat i\hat j}=\mathrm{diag}(-1,1,1)$. In this basis, spherical symmetry implies that the two angular stresses are equal. Moreover, there is no independent radial surface pressure, because the radial direction is normal to the shell and does not belong to the intrinsic geometry of $\Sigma$. Thus, the surface stress--energy tensor can be written as
\begin{equation}
	S_{\hat i\hat j}=\mathrm{diag}(\sig,p,p),
	\label{eq:surface_tensor}
\end{equation}
where $\sig$ is the surface energy density and $p$ is the tangential pressure on the shell.

The surface stresses are determined by the Lanczos junction equation, obtained from the jump of the extrinsic curvature across the shell~\cite{Darmois1927,Lanczos1924,Israel1966},
\begin{equation}
	-[K_{\hat i\hat j}]+[K]h_{\hat i\hat j}=8\pi S_{\hat i\hat j},
	\qquad
	[K_{\hat i\hat j}]=K^+_{\hat i\hat j}-K^-_{\hat i\hat j},
	\qquad
	[K]=h^{\hat i\hat j}[K_{\hat i\hat j}].
	\label{eq:lanczos}
\end{equation}
The quantity $[K_{\hat i\hat j}]$ measures the discontinuity of the extrinsic curvature at the junction. If this jump vanishes, the two copies are smoothly matched at the level of the first normal derivative of the metric, and no distributional surface stress tensor is generated. If the jump is nonzero, the metric remains continuous but its normal derivative changes across the shell. This produces a surface layer whose stress tensor is precisely $S_{\hat i\hat j}$. In the present symmetric cut-and-paste construction, this surface layer is the exotic matter supporting the thin-shell wormhole.

Using Eqs.~\eqref{eq:K_tautau} and \eqref{eq:K_angular}, the dynamical surface energy density and tangential pressure are
\begin{equation}
	\sig(\chi)=-\frac{1}{2\pi\chi}\sqrt{\Gv(\chi)+\dot\chi^2},
	\label{eq:lambda_dynamic}
\end{equation}
\begin{equation}
	p(\chi)=\frac{1}{8\pi}\left[
	\frac{\Gv'(\chi)+2\ddot\chi}{\sqrt{\Gv(\chi)+\dot\chi^2}}
	+\frac{2\sqrt{\Gv(\chi)+\dot\chi^2}}{\chi}
	\right].
	\label{eq:pressure_dynamic}
\end{equation}
For a static throat, $\chi=\chi_0$ with $\dot\chi=\ddot\chi=0$, these expressions reduce to
\begin{equation}
	\sig_0=-\frac{\sqrt{G_{{\rm v}0}}}{2\pi\chi_0},
	\label{eq:lambda_static}
\end{equation}
\begin{equation}
	p_0=\frac{\chi_0G_{{\rm v}0}'+2G_{{\rm v}0}}{8\pi\chi_0\sqrt{G_{{\rm v}0}}},
	\label{eq:pressure_static}
\end{equation}
where $G_{{\rm v}0}=\Gv(\chi_0)$ and $G_{{\rm v}0}'=\Gv'(\chi_0)$. Since the throat must lie in the positive-lapse region, $G_{{\rm v}0}>0$, Eq.~\eqref{eq:lambda_static} shows that the surface energy density is necessarily negative. This is a direct consequence of the symmetric thin-shell construction: the flare-out is produced by the jump in the angular extrinsic curvature, and the corresponding surface density is negative.

\begin{figure}[htp!]
	\centering
	\includegraphics[width=0.49\linewidth]{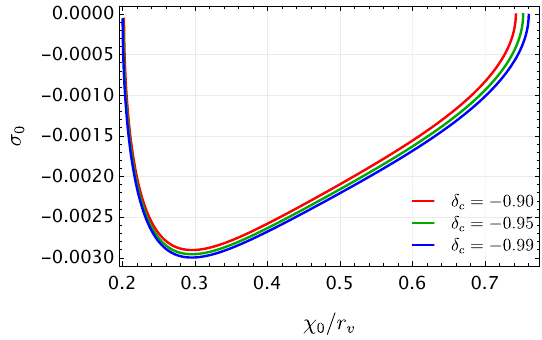}
	\includegraphics[width=0.49\linewidth]{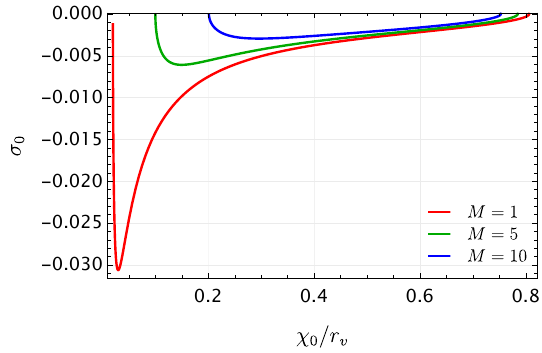}
	\caption{Static surface energy density $\sig_0$ as a function of the dimensionless throat radius $\chi_0/r_v$. In the left panel, the mass is fixed at $M=10$ and the curves correspond to $\delta_c=-0.90$, $-0.95$, and $-0.99$. In the right panel, the density contrast is fixed at $\delta_c=-0.95$ and the curves correspond to $M=1$, $5$, and $10$. In both panels, we use $r_s=80$, $r_v=100$, $\rho_0=3\times10^{-5}$, $\alpha=2.4$, and $\beta=7.5$. The curves are shown only in the positive-lapse region between the black-hole horizon and the cosmological-like horizon, where the static thin-shell construction is well defined.}
	\label{fig:sigma0_static}
\end{figure}

Figure~\ref{fig:sigma0_static} complements the analytical result in Eq.~\eqref{eq:lambda_static} by showing how the amount of negative surface energy varies across the admissible static interval. In both panels, $\sig_0$ approaches zero near the boundaries of the plotted domain and develops a negative minimum at an intermediate throat radius. This behavior reflects the fact that the shell is being moved between two horizon-like boundaries of the static patch: near either boundary the lapse tends to zero, while in the interior it reaches larger positive values and increases the magnitude of the surface density.

The left panel shows that, for fixed $M=10$, the dependence on the void depth is relatively mild. The three curves are close over most of the interval, indicating that the surface density near the inner part of the static region is mainly controlled by the central mass. The effect of $\delta_c$ becomes more visible toward the outer region, where deeper voids slightly enlarge the allowed interval and shift the return of $\sig_0$ toward zero to larger values of $\chi_0/r_v$.

The right panel shows a stronger sensitivity to the central mass. Smaller values of $M$ allow the shell to be placed at smaller dimensionless radii, where the geometrical prefactor in the surface density makes the negative minimum considerably deeper. Conversely, increasing $M$ shifts the inner boundary outward and suppresses the magnitude of $\sig_0$ throughout the allowed region. Thus, although the sign of the surface density is fixed by the symmetric junction, the amount of negative surface energy required to support the throat is strongly affected by the position of the black-hole horizon.

This behavior should not be interpreted as a complete weakening of exoticity near the cosmological-like horizon. It only shows that the surface density itself tends to zero in that limit. The full energy-condition analysis still depends on the tangential pressure, and therefore on the combinations examined below.

\subsection{Energy conditions on the shell}

The energy conditions on the shell must be interpreted intrinsically, because $S_{\hat i\hat j}$ is a surface tensor rather than a bulk stress tensor. The null energy condition (NEC) requires
\begin{equation}
	\sig_0+p_0\geq0.
	\label{eq:nec_condition}
\end{equation}
The weak energy condition (WEC) requires
\begin{equation}
	\sig_0\geq0,
	\qquad
	\sig_0+p_0\geq0,
	\label{eq:wec_condition}
\end{equation}
while the dominant energy condition (DEC) demands, in addition to non-negative energy density, that the magnitude of the tangential pressure does not exceed the energy density, $|p_0|\leq -\sig_0$. For the present static symmetric shell, Eq.~\eqref{eq:lambda_static} already implies $\sig_0<0$, and therefore both the WEC and DEC are necessarily violated at the throat. The strong energy condition (SEC) on the intrinsic shell is controlled by
\begin{equation}
	\sig_0+p_0\geq0,
	\qquad
	\sig_0+2p_0\geq0.
	\label{eq:sec_condition}
\end{equation}

This situation differs from the Morris--Thorne wormhole in an important way. In a Morris--Thorne geometry, the exotic matter is usually described by a bulk stress tensor, with an energy density, a radial pressure, and tangential pressures distributed in a finite region around the throat. There the NEC violation typically appears as a bulk combination $\rho+p_r<0$ near the throat. Here, by contrast, the bulk is fixed by the void geometry on each side, and the exotic matter is localized on the junction hypersurface. Hence the relevant energy-condition combinations are not bulk combinations such as $\rho+p_r$, but intrinsic surface combinations such as $\sig+p$.

For the static shell, the null-energy combination is
\begin{equation}
	\sig_0+p_0=
	\frac{\chi_0G_{{\rm v}0}'-2G_{{\rm v}0}}{8\pi\chi_0\sqrt{G_{{\rm v}0}}}.
	\label{eq:nec_static}
\end{equation}
The trace-type combination entering the intrinsic strong energy condition is
\begin{equation}
	\sig_0+2p_0=
	\frac{G_{{\rm v}0}'}{4\pi\sqrt{G_{{\rm v}0}}}.
	\label{eq:sec_static}
\end{equation}
Using $G_{{\rm v}0}=1-2m_{{\rm v}0}/\chi_0$ and $G_{{\rm v}0}'=2m_{{\rm v}0}/\chi_0^{2}-8\pi\chi_0\rho_{{\rm v}0}$, these combinations can be expressed directly in terms of the void mass function and the void density profile. Therefore, although the negative sign of $\sig_0$ enforces the violation of the weak and dominant energy conditions, the signs of $\sig_0+p_0$ and $\sig_0+2p_0$ depend on the local behavior of the lapse function at the throat. In the void geometry, this means that the NEC and SEC combinations are controlled by the position of the shell inside the positive-lapse interval between the black-hole horizon and the cosmological-like horizon.

\begin{figure}[htp!]
	\centering
	\includegraphics[width=0.49\linewidth]{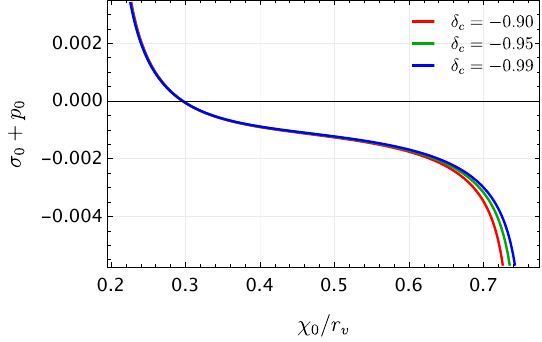}
	\includegraphics[width=0.49\linewidth]{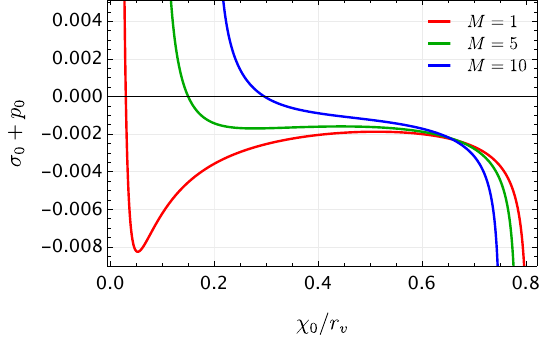}
	\includegraphics[width=0.49\linewidth]{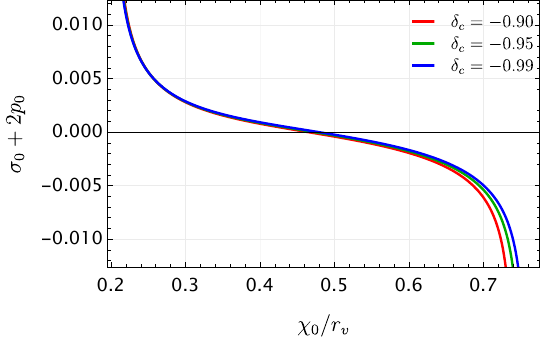}
	\includegraphics[width=0.49\linewidth]{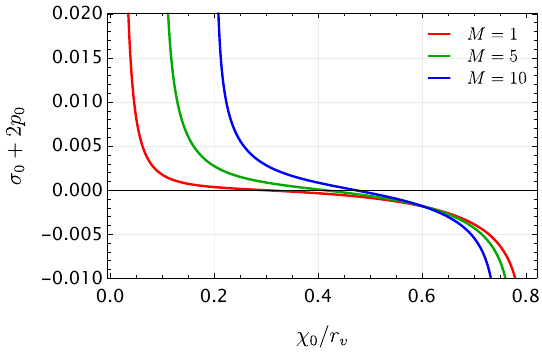}
	\caption{Static energy-condition combinations as functions of the dimensionless throat radius $\chi_0/r_v$. The upper panels show the NEC combination $\sig_0+p_0$, while the lower panels show the trace-type combination $\sig_0+2p_0$ entering the intrinsic SEC. In the left column, the mass is fixed at $M=10$ and the curves correspond to $\delta_c=-0.90$, $-0.95$, and $-0.99$. In the right column, the density contrast is fixed at $\delta_c=-0.95$ and the curves correspond to $M=1$, $5$, and $10$. In all panels, we use $r_s=80$, $r_v=100$, $\rho_0=3\times10^{-5}$, $\alpha=2.4$, and $\beta=7.5$. The horizontal line marks the zero level separating satisfied and violated energy-condition combinations.}
	\label{fig:nec_sec_static}
\end{figure}

Figure~\ref{fig:nec_sec_static} shows that the energy-condition behavior of the shell is not determined by the surface density alone. The upper panels indicate that the NEC combination is positive only near the inner black-hole horizon and becomes negative throughout most of the admissible interval. This change of sign marks the transition from a region where the tangential pressure compensates the negative surface density to a region where the pressure contribution reinforces the exotic character of the shell. Toward the cosmological-like horizon, the NEC violation becomes increasingly strong, showing that the apparent reduction of $|\sig_0|$ in Fig.~\ref{fig:sigma0_static} does not correspond to a regular matter-free limit.

This point is physically important. The limit $\sig_0\to0^-$ near the outer boundary should not be interpreted as the disappearance of the material content of the shell. The surface layer is described by the full tensor $S_{\hat i\hat j}$, and therefore the tangential pressure also contributes to its physical character. As the throat approaches the cosmological-like horizon from inside the static region, the surface density tends to zero, but the pressure contribution becomes increasingly negative. Hence, the shell approaches a tension-dominated regime rather than a vacuum one. In this sense, configurations very close to $r_{++}$ are not regular low-exoticity throats, but horizon-limit configurations supported by extreme tangential stresses.

The same qualitative tendency is observed in the SEC trace combination shown in the lower panels. Near the inner boundary, the combination is positive, but it decreases as the throat is moved outward and eventually becomes negative before the shell approaches the outer horizon. Therefore, the outer part of the static patch is not only associated with negative surface energy density, but also with increasingly restrictive tangential stresses. This behavior is a direct consequence of placing a static shell in a geometry with two horizon-like boundaries \cite{SharifMumtaz2014}: the inner and outer limits of the admissible interval are both characterized by a vanishing lapse, but the sign of the radial derivative of the lapse changes the role played by the pressure term.

The dependence on the void depth remains relatively mild when $M$ is fixed. As in the surface-density plot, the curves associated with different values of $\delta_c$ are close over most of the interval, with the most visible differences occurring near the outer region. Deeper voids slightly shift the cosmological-like horizon outward and consequently extend the interval in which the shell can be placed. However, they do not qualitatively change the energy-condition pattern: the NEC and the SEC trace combination are satisfied only in the inner portion of the static patch and are violated as the throat approaches the cosmological-like horizon.

The dependence on the central mass is more pronounced. Increasing $M$ displaces the black-hole horizon to larger values of $\chi_0/r_v$, changing the size of the inner region where the NEC and the SEC trace combination are positive. The curves also show that configurations with smaller masses can probe smaller dimensionless radii, where the surface quantities vary more sharply. In contrast, larger masses push the admissible throat outward and modify the balance between surface density and tangential pressure. Thus, the central mass controls not only the available static interval, but also the location at which the shell changes from a partially energy-condition-satisfying regime to a clearly exotic one.

\section{Thermodynamics of the thin‑shell wormhole}
\label{sec:thermo}

	The static thin‑shell configuration constructed in the previous sections can be assigned thermodynamic properties that connect the void geometry, the shell matter, and the two horizon‑like boundaries of the seed spacetime. In this section we define the temperatures of the relevant geometric objects, examine the first law for the shell, and discuss how the void environment enters the thermodynamic description.
	
	\subsection{Temperatures, first law, and mass variation}
	
	The void metric~\eqref{eq:void_metric} possesses two positive roots of the lapse function: the black‑hole horizon \(r_h\) and the cosmological‑like horizon \(r_{++}\). In a static patch with a non‑degenerate Killing horizon, the surface gravity \(\kappa\) is obtained from the derivative of the lapse function,
	\begin{equation}
		\kappa_{\mathcal H}=\frac12\Bigl|G_{\rm v}'(r_{\mathcal H})\Bigr|,
		\label{eq:surface_gravity}
	\end{equation}
	and the associated Hawking temperature is \(T_{\mathcal H}=\kappa_{\mathcal H}/2\pi\). For the two horizons one obtains
	\begin{equation}
		T_h=\frac{G_{\rm v}'(r_h)}{4\pi},\qquad
		T_{++}=-\frac{G_{\rm v}'(r_{++})}{4\pi},
		\label{eq:horizon_temperatures}
	\end{equation}
	where the minus sign in \(T_{++}\) ensures positivity because \(G_{\rm v}'(r_{++})<0\). Their Bekenstein–Hawking entropies are simply one quarter of the horizon areas,
	\begin{equation}
		S_h=\pi r_h^{2},\qquad
		S_{++}=\pi r_{++}^{2},
		\label{eq:horizon_entropies}
	\end{equation}
	in geometrized units \(G=c=1\).
	
	The shell itself is not a horizon, but an observer comoving with a static shell at \(r=\chi\) experiences a proper acceleration whose magnitude is
	\begin{equation}
		a=\frac{|G_{\rm v}'(\chi)|}{2\sqrt{G_{\rm v}(\chi)}}.
		\label{eq:proper_acceleration}
	\end{equation}
	Through the Unruh effect, this acceleration defines a local temperature
	\begin{equation}
		T_{\rm shell}=\frac{a}{2\pi}= \frac{|G_{\rm v}'(\chi)|}{4\pi\sqrt{G_{\rm v}(\chi)}}.
		\label{eq:shell_temperature}
	\end{equation}
	Equation~\eqref{eq:shell_temperature} is the temperature measured by a static observer at the throat and is the natural candidate for the thermodynamic temperature of the shell fluid. This identification of the shell temperature with the local Unruh temperature has recently been extended to dynamical thin‑shell wormholes in~\cite{Lobo:2026qhw}, where a generalized Unruh identity equates the acceleration temperature to a Hawking‑like particle‑creation temperature governed by the peeling of null rays. The absolute value guarantees positivity across the whole static patch, regardless of the sign of \(G_{\rm v}'\).
	
	The shell carries a surface energy density \(\sigma\) and a tangential pressure \(p\). The total internal energy of the shell is \(E=\sigma A\), where \(A=4\pi\chi^{2}\) is the throat area. Treating the shell as a closed, two‑dimensional fluid, we adopt the standard thermodynamic first law
	\begin{equation}
		T_{\rm shell}\,\dd S_{\rm shell}= \dd E + p\,\dd A,
		\label{eq:first_law_shell}
	\end{equation}
	where \(S_{\rm shell}\) is the entropy of the matter on the throat. This framework is consistent with the unified thermodynamic description of thin‑shell wormholes developed in~\cite{Lobo:2026vrn}, where it was shown that for isolated shells the entropy is conserved under transparent evolution, i.e.\ \(T\dd S=0\), and that a generalized first law emerges when bulk matter crosses the throat. For a static configuration at a fixed central mass \(M\), the expressions~\eqref{eq:lambda_static} and~\eqref{eq:pressure_static} imply
	\begin{equation}
		E = \sigma_0 A = -2\chi\sqrt{G_{\rm v}(\chi)},
		\qquad
		p_0\,\frac{\dd A}{\dd\chi}= \frac{\chi G_{\rm v}'+2G_{\rm v}}{\sqrt{G_{\rm v}}}.
		\label{eq:E_and_pdAdchi}
	\end{equation}
	A direct calculation yields
	\begin{equation}
		\frac{\dd E}{\dd\chi}+p_0\frac{\dd A}{\dd\chi}=0,
		\label{eq:static_first_law_vanishes}
	\end{equation}
	so that for variations that keep the mass fixed the first law~\eqref{eq:first_law_shell} reduces to
	\begin{equation}
		T_{\rm shell}\frac{\dd S_{\rm shell}}{\dd\chi}=0.
		\label{eq:S_shell_constant}
	\end{equation}
	Thus, along a sequence of static throats with the same \(M\) the entropy of the shell does not change; it can be set to a constant, for instance to zero. This is in agreement with the general result of~\cite{Lobo:2026vrn}, where the shell entropy was found to be conserved for isolated configurations in the absence of bulk fluxes. Physically, this reflects that the work done by the pressure exactly compensates the change in internal energy, leaving no room for heat flow.
	
	When the ADM mass \(M\) is allowed to vary, the internal energy acquires an explicit dependence on \(M\) through \(G_{\rm v}\). From $E=-2\chi\sqrt{G_{\rm v}(\chi,M)}$ and $\partial G_{\rm v}/\partial M = -2/\chi$, one finds
	\begin{equation}
		\left(\frac{\partial E}{\partial M}\right)_\chi
		= -\frac{\chi}{\sqrt{G_{\rm v}}}\frac{\partial G_{\rm v}}{\partial M}
		= \frac{2}{\sqrt{G_{\rm v}}}.
		\label{eq:dEdM}
	\end{equation}
	Because the equilibrium condition $dE + p\,dA = 0$ holds for fixed $M$, the only change in the shell's internal energy along a family of static solutions comes from the variation of the mass. The first law~\eqref{eq:first_law_shell} therefore becomes
	\begin{equation}
		T_{\rm shell}\,\dd S_{\rm shell}
		= \frac{2}{\sqrt{G_{\rm v}}}\,\dd M.
		\label{eq:first_law_with_M}
	\end{equation}
	Using the expression~\eqref{eq:shell_temperature} for the temperature, one obtains
	\begin{equation}
		\dd S_{\rm shell}= \frac{8\pi}{|G_{\rm v}'(\chi)|}\,\dd M.
		\label{eq:dS_shell_vs_dM}
	\end{equation}
	Equation~\eqref{eq:dS_shell_vs_dM} shows that the entropy of the shell can be expressed as an integral over the mass once an equilibrium relation \(\chi(M)\) is provided, e.g.\ by an equation of state. For the Chaplygin‑type models studied in Sec.~\ref{sec:gccg}–\ref{sec:mccg}, the static junction condition fixes \(B\) in terms of \(\chi\) and \(M\), and the stability analysis restricts the allowed throat positions. In such cases, \(\chi\) is determined by \(M\), and Eq.~\eqref{eq:dS_shell_vs_dM} can be integrated to yield a finite shell entropy.
	
\subsection{Horizon thermodynamics and entropy relations}

When only the mass \(M\) is varied and the void parameters \((\rho_0,\delta_c,r_s,r_v)\) are kept fixed, the two horizon radii become functions of \(M\) alone. From the horizon condition \(G_{\rm v}(r_{\mathcal H},M)=0\) one obtains
\begin{equation}
	G_{\rm v}'(r_{\mathcal H})\,\dd r_{\mathcal H} - \frac{2}{r_{\mathcal H}}\,\dd M =0,
	\label{eq:horizon_differential}
\end{equation}
which, together with the definitions~\eqref{eq:horizon_temperatures} and~\eqref{eq:horizon_entropies}, implies
\begin{equation}
	\dd M = T_h\,\dd S_h, \qquad
	\dd M = -\,T_{++}\,\dd S_{++}.
	\label{eq:horizon_first_laws}
\end{equation}
The opposite signs reflect the fact that the black‑hole horizon grows with \(M\) while the cosmological‑like horizon shrinks; the two horizon entropies are not independent but satisfy the constraint
\[
T_h\,\dd S_h + T_{++}\,\dd S_{++}=0.
\]
Inserting the first relation of~\eqref{eq:horizon_first_laws} into~\eqref{eq:dS_shell_vs_dM} gives a direct connection between the shell entropy and the black‑hole entropy:
\begin{equation}
	\dd S_{\rm shell}= \frac{8\pi}{|G_{\rm v}'(\chi)|}\, T_h\,\dd S_h.
	\label{eq:shell_vs_bh_entropy}
\end{equation}
In the limit where the throat approaches the black‑hole horizon, \(\chi\to r_h\), one has \(G_{\rm v}(\chi)\to0\) and \(|G_{\rm v}'(\chi)|\to G_{\rm v}'(r_h)=4\pi T_h\). The right‑hand side of~\eqref{eq:shell_vs_bh_entropy} then reduces to \(2\,\dd S_h\), indicating that the shell entropy remains finite and of the same order as the black‑hole entropy in that regime. If the throat approaches the cosmological‑like horizon, the analogous relation \(\dd S_{\rm shell}= -\,(8\pi/|G_{\rm v}'(\chi)|) T_{++}\,\dd S_{++}\) gives \(\dd S_{\rm shell}\to -2\,\dd S_{++}\).

When the central mass is negligible compared with the void contribution, the lapse is approximately de Sitter‑like, \(G_{\rm v}\simeq 1-\Lambda_{\rm v}r^{2}/3\) with \(\Lambda_{\rm v}=8\pi\rho_0(1+\delta_c)\). In this limit the horizon radii become \(r_h\simeq 0\) and \(r_{++}\simeq \sqrt{3/\Lambda_{\rm v}}\), while the static throat radius is constrained only by the de Sitter horizon. From Eq.~\eqref{eq:dS_shell_vs_dM} one has \(|G_{\rm v}'|\simeq 2\Lambda_{\rm v}\chi/3\) and therefore
\begin{equation}
	\dd S_{\rm shell}\simeq \frac{12\pi}{\Lambda_{\rm v}\chi}\,\dd M.
	\label{eq:dS_shell_desitter}
\end{equation}
If, in addition, the throat radius is fixed by a barotropic equation of state, \(\chi\) becomes a function of \(M\) and the integration can be performed. For instance, for a fluid with \(p=w\sigma\) the static junction condition forces \(\chi\) to scale as \(M^{1/(3w+1)}\), leading to a power‑law behaviour of \(S_{\rm shell}(M)\). More generally, integrating Eq.~\eqref{eq:dS_shell_vs_dM} yields
\begin{equation}
	S_{\rm shell}(M)=8\pi\int_{M_{\rm ref}}^{M}\frac{\dd M'}{|G_{\rm v}'(\chi(M'))|}.
	\label{eq:Shell_entropy_integrated}
\end{equation}

These results show that the shell entropy is directly tied to the horizon entropies through the mass variation, and that the precise form of a Smarr‑type relation will depend on the specific equation of state and the function \(\chi(M)\).  While the shell behaves as an additional thermodynamic component coupled to the horizons, a closed integrated Smarr formula analogous to the familiar black‑hole relation would require specifying the exact equilibrium sequence, which is left for future investigation.
	
\subsection{Comparison with Schwarzschild thin‑shell wormholes}

When the void density is completely removed (\(\rho_0\to0\), hence \(G_{\rm v}\to1-2M/r\)), the cosmological‑like horizon disappears and the static patch extends to infinity. In that limit, the surface energy density and pressure reduce to the well‑known Schwarzschild thin‑shell expressions~\cite{PoissonVisser1995}, and the shell temperature becomes \(T_{\rm shell}^{\rm (Sch)}= M/(2\pi\chi^{2}\sqrt{1-2M/\chi})\). The first law yields \(\dd S_{\rm shell}^{\rm (Sch)}= 4\pi\chi^{2}/M\,\dd M\), which is qualitatively different from the void case because the denominator no longer involves a derivative but the mass itself. The presence of the void thus introduces an external scale (\(\Lambda_{\rm v}\)) that enriches the thermodynamic phase space, allowing finite‑temperature shells even in regions where the Schwarzschild temperature would vanish at infinity.

Thus, the thin‑shell wormhole inherits a nontrivial thermodynamic structure from the void environment. The static shell has no entropy variation at fixed mass, but once the mass is treated as a thermodynamic variable, a first law emerges that connects the shell entropy to the black‑hole (or cosmological‑like) horizon entropy. The explicit form of \(S_{\rm shell}\) depends on the equation of state and on the relation \(\chi(M)\) dictated by the stability conditions. This analysis aligns with the broader thermodynamic framework for thin‑shell wormholes presented in~\cite{Lobo:2026vrn}, where the generalised first law and second law were established for dynamic configurations with bulk fluxes. The present void‑embedded wormhole can be seen as a concrete realization of that framework in a cosmological setting. For the GCCG and MCCG models discussed in this work, the stability analysis effectively selects a particular static configuration for each \(M\), allowing \(S_{\rm shell}\) to be computed as a function of the black‑hole mass and the void parameters. This reinforces the interpretation of the thin‑shell wormhole as a composite object whose global thermodynamic properties are controlled by the interplay between the local exotic matter and the large‑scale de Sitter‑like environment.

\section{Conservation equation and radial dynamics}\label{sec:dynamics}

\subsection{Conservation equation and effective potential}

The algebraic junction equation determines the instantaneous surface stresses once the shell trajectory is specified. To describe the radial evolution of the throat, however, one must also impose the intrinsic conservation law on the hypersurface. This relation follows from the Codazzi--Mainardi equation and, following the notation commonly used in thin-shell analyses~\cite{Javed2024}, can be written as
\begin{equation}
	-\nabla_i S^i{}_j
	=
	\left[
	T_{\alpha\beta}
	\frac{\partial x^\alpha}{\partial \xi^j}
	\eta^\beta
	\right]
	\equiv \mathcal{F}_j,
	\label{eq:lanczos_covariant_conservation}
\end{equation}
where $\nabla_i$ denotes the covariant derivative compatible with the induced metric on $\Sigma$. The vector $\mathcal{F}_j$ represents the net flux of energy-momentum from the bulk into the shell. Therefore, Eq.~\eqref{eq:lanczos_covariant_conservation} states that the surface energy-momentum is conserved intrinsically only when there is no net exchange with the surrounding spacetime.

For the radial dynamics, the relevant component is the projection along the shell four-velocity $u^i$, namely $u^j\mathcal{F}_j \equiv \mathcal{F}$. In differential form, the conservation equation projected along $u^i$ becomes
\begin{equation}
	\frac{\dd}{\dd\tau}\left(4\pi\chi^2\sig\right)
	+p\frac{\dd}{\dd\tau}\left(4\pi\chi^2\right)
	=
	4\pi\chi^2\mathcal{F},
	\label{eq:conservation_tau_flux}
\end{equation}
where $\mathcal{F}$ is the energy flux measured by an observer comoving with the shell. The first term gives the change of the total internal energy of the shell, while the second term is the work done by the tangential pressure during the change of the shell area. For the void metric~\eqref{eq:void_metric}, the bulk stress-energy tensor satisfies $T^{t}{}_{t}=T^{r}{}_{r}$, thus the normal component of the bulk flux then vanishes, $\mathcal{F}=0$~\cite{Garcia2012}, because $\eta^\mu$ is purely radial while the only non‑zero off‑diagonal components of $T_{\mu\nu}$ would be $T_{tr}$, which are absent. A nonzero right-hand side would represent an external exchange of energy or momentum between the shell and the bulk geometry, modifying the intrinsic conservation of the surface fluid.

Thus, in the present case one obtains
\begin{equation}
	\frac{\dd}{\dd\tau}\left(4\pi\chi^2\sig\right)
	+p\frac{\dd}{\dd\tau}\left(4\pi\chi^2\right)=0,
	\label{eq:conservation_tau}
\end{equation}
or, equivalently,
\begin{equation}
	\dot\sig+2\frac{\dot\chi}{\chi}(\sig+p)=0.
	\label{eq:conservation_tau_local}
\end{equation}
Whenever $\dot\chi\neq0$, this can be rewritten as a radial equation,
\begin{equation}
	\sig'(\chi)=-\frac{2}{\chi}\left[\sig(\chi)+p(\chi)\right].
	\label{eq:conservation_radial}
\end{equation}

Equation~\eqref{eq:conservation_radial} is not closed until the matter content of the shell is specified. An equation of state provides precisely this missing information. In the simplest barotropic case, one assumes $p=p(\sig)$, while a more general surface fluid may have an explicit dependence on the throat radius, $p=p(\chi,\sig)$. Once such a relation is chosen, Eq.~\eqref{eq:conservation_radial} determines $\sig(\chi)$ from a given initial or static value. In this way, the surface density becomes a function of the throat position, and the radial motion can be written as a one-dimensional effective problem.

Solving the dynamical junction condition in Eq.~\eqref{eq:lambda_dynamic} for $\dot\chi^2$ gives the equation of motion
\begin{equation}
	\dot\chi^2+\Ueff(\chi)=0,
	\label{eq:eom_effective_potential}
\end{equation}
with the effective potential
\begin{equation}
	\Ueff(\chi)=\Gv(\chi)-4\pi^2\chi^2\sig^2(\chi).
	\label{eq:effective_potential_general}
\end{equation}
Equation~\eqref{eq:eom_effective_potential} is formally equivalent to the energy conservation of a classical particle of unit mass in a potential $\Ueff(\chi)$. The first term is purely geometrical and comes from the void lapse function, whereas the second term is determined by the surface matter distribution. With the sign convention adopted in Eq.~\eqref{eq:eom_effective_potential}, the kinetic term satisfies $\dot\chi^2=-\Ueff(\chi)$, so the admissible motion is restricted to regions where $\Ueff(\chi)\leq0$. Points where $\Ueff(\chi)=0$ correspond to radii at which the shell is instantaneously at rest and may represent turning points of the motion. A genuine static equilibrium requires, in addition, that $\Ueff'(\chi_0)=0$.

A static configuration at $\chi=\chi_0$ corresponds to an equilibrium point of the effective potential. Therefore,
\begin{equation}
	\Ueff(\chi_0)=0,
	\qquad
	\Ueff'(\chi_0)=0.
	\label{eq:static_equilibrium_conditions}
\end{equation}
Expanding the potential around this radius gives
\begin{equation}
	\Ueff(\chi)=
	\frac{1}{2}\Ueff''(\chi_0)(\chi-\chi_0)^2
	+\mathcal{O}\left[(\chi-\chi_0)^3\right],
	\label{eq:potential_expansion}
\end{equation}
where the zeroth- and first-order terms vanish by Eq.~\eqref{eq:static_equilibrium_conditions}. Hence the sign of $\Ueff''(\chi_0)$ determines the local response of the shell to radial perturbations. If
\begin{equation}
	\Ueff''(\chi_0)>0,
	\label{eq:stability_condition}
\end{equation}
the potential is locally positive away from the equilibrium point, so nearby radial motion is forbidden by Eq.~\eqref{eq:eom_effective_potential}; the static throat is therefore linearly stable. If $\Ueff''(\chi_0)<0$, the potential becomes negative in a neighborhood of $\chi_0$, allowing the throat to move away from equilibrium, which signals instability.

For a general surface equation of state, define
\begin{equation}
	\zeta(\chi)=\left(\frac{\partial p}{\partial\sig}\right)_\chi,
	\qquad
	p_\chi(\chi)=\left(\frac{\partial p}{\partial\chi}\right)_\sig.
	\label{eq:zeta_definition}
\end{equation}
Differentiating Eq.~\eqref{eq:effective_potential_general} and using the radial conservation law, the second derivative of the potential can be written as
\begin{equation}
	\begin{aligned}
		\Ueff''(\chi)=\Gv''(\chi)
		&-8\pi^2\left\{[\sig(\chi)+2p(\chi)]^2
		+2\sig(\chi)[\sig(\chi)+p(\chi)]\left[1+2\zeta(\chi)\right]\right\}+16\pi^2\chi\,\sig(\chi)\,p_\chi(\chi).
	\end{aligned}
	\label{eq:Vpp_general_eos}
\end{equation}
For a barotropic surface fluid, $p=p(\sig)$, the explicit-radius term vanishes, $p_\chi=0$. This expression shows that the stability of the throat is controlled by two sectors: the local curvature of the void lapse function and the response of the surface pressure to changes in the surface density. The stability problem is therefore reduced to evaluating Eq.~\eqref{eq:Vpp_general_eos} at the static radius after the equation of state has fixed the function $\sig(\chi)$.

For the numerical implementation, it is important to distinguish the infinitesimal displacement used in the linear stability analysis from the parameter employed to scan different static throat locations. The dynamical perturbation corresponds to a small time-dependent displacement of the shell around a fixed equilibrium radius $\chi_0$, whereas the quantity $\mathcal{E}$ used in the numerical plots is not this perturbation. Instead, it is a radial coordinate offset that labels a family of static throats located near the inner horizon, according to
\begin{equation}
	\chi_0=r_h+|\mathcal{E}|.
	\label{eq:horizon_offset_parameterization}
\end{equation}
This parametrization is designed to probe the stability properties of configurations that are increasingly close to the black‑hole horizon, where the strongest gravitational effects occur. For each value of $\mathcal{E}$, the static surface quantities and any equation-of-state parameter fixed by the junction conditions are recalculated, and the corresponding value of $\Ueff''(\chi_0)$ is evaluated. Only radii satisfying $r_h<\chi_0<\min(r_{++},r_v)$ and $\Gv(\chi_0)>0$ are retained. Therefore, varying $\mathcal{E}$ scans different equilibrium configurations near the black-hole horizon, while the linear stability criterion refers to infinitesimal radial displacements around each selected static throat. Since the parametrization involves $|\mathcal{E}|$, positive and negative values of $\mathcal{E}$ correspond to the same static radius, and the resulting stability curves are symmetric by construction.

\subsection{Generalized cosmic Chaplygin gas (GCCG)}\label{sec:gccg}

For the GCCG~\cite{GonzalezDiaz2003,SharifMumtaz2014}, the surface equation of state is taken as
\begin{equation}
	p^{\rm (G)}(\chi)=-\frac{1}{\left[\sig^{\rm (G)}(\chi)\right]^\gamma}
	\left[E^{\rm (G)}+
	\left(\left[\sig^{\rm (G)}(\chi)\right]^{1+\gamma}-E^{\rm (G)}\right)^{-\omega}\right],
	\label{eq:gccg_eos}
\end{equation}
where
\begin{equation}
	E^{\rm (G)}=\frac{B^{\rm (G)}}{1+\omega}-1,
	\qquad
	B^{\rm (G)}\in(-\infty,\infty),
	\qquad
	-C<\omega<0,
	\label{eq:gccg_E_definition}
\end{equation}
with $C>0$ and $C\neq1$. The parameter $\gamma$ is the generalized Chaplygin index and is usually taken in the range $0<\gamma\leq1$ \cite{SharifMumtaz2014}. It controls the nonlinear dependence of the pressure on the surface density: smaller values of $\gamma$ produce a weaker inverse-density response, whereas $\gamma=1$ gives the ordinary Chaplygin-type inverse-density behavior. The parameter $B^{\rm (G)}$ fixes the amplitude of the Chaplygin sector, while $E^{\rm (G)}$ is a convenient reparametrization of $B^{\rm (G)}$ adapted to the GCCG form. The parameter $\omega$ controls the cosmic Chaplygin deformation; it measures the departure from the generalized Chaplygin gas and is restricted to negative values in the GCCG model.

The standard Chaplygin-type equations of state are recovered as limiting cases. In the limit $\omega\to0$, Eq.~\eqref{eq:gccg_eos} reduces to the generalized Chaplygin gas. If, in addition, $\gamma=1$, one obtains the usual Chaplygin gas,
\begin{equation}
	p^{\rm (G)}(\chi)\longrightarrow
	-\frac{B^{\rm (G)}}{\sig^{\rm (G)}(\chi)}.
	\label{eq:gccg_cg_limit}
\end{equation}
Thus, the GCCG contains the generalized Chaplygin gas and the standard Chaplygin gas as particular limiting cases, while the parameter $\omega$ introduces a further nonlinear correction to the pressure-density relation. The radial conservation equation \eqref{eq:conservation_radial} becomes
\begin{equation}
	\frac{\dd\sig^{\rm (G)}}{\dd\chi}
	=-\frac{2}{\chi}\left\{\sig^{\rm (G)}
	-\frac{1}{\left(\sig^{\rm (G)}\right)^\gamma}
	\left[E^{\rm (G)}+
	\left(\left(\sig^{\rm (G)}\right)^{1+\gamma}-E^{\rm (G)}\right)^{-\omega}\right]\right\}.
	\label{eq:gccg_conservation}
\end{equation}
Thus, the GCCG effective potential is
\begin{equation}
	\Ueff^{\rm (G)}(\chi)=\Gv(\chi)-4\pi^2\chi^2\left[\sig^{\rm (G)}(\chi)\right]^2.
	\label{eq:gccg_potential}
\end{equation}

The static value of the shell density is fixed geometrically by Eq.~\eqref{eq:lambda_static}. The Chaplygin parameter $B^{\rm (G)}$ is fixed by requiring the GCCG pressure to reproduce the static junction pressure in Eq.~\eqref{eq:pressure_static}. Equivalently,
\begin{equation}
	p_0+\frac{1}{\sig_0^\gamma}
	\left[E^{\rm (G)}_0+
	\left(\sig_0^{1+\gamma}-E^{\rm (G)}_0\right)^{-\omega}\right]=0,
	\label{eq:gccg_static_E_condition}
\end{equation}
where
\begin{equation}
	B^{\rm (G)}_0=(1+\omega)\left(E^{\rm (G)}_0+1\right).
	\label{eq:gccg_B_static}
\end{equation}
Therefore $B^{\rm (G)}$ is not an independent parameter once $\chi_0$, the void parameters, $\gamma$, and $\omega$ are chosen. Consequently, the stability of the GCCG configuration is entirely determined by the void geometry and the remaining equation‑of‑state parameters $(\gamma,\omega)$. The real branch of Eq.~\eqref{eq:gccg_static_E_condition} must be selected consistently with the powers appearing in Eq.~\eqref{eq:gccg_eos}.

For the GCCG model,
\begin{equation}
	\zeta^{\rm (G)}(\chi)=\omega(1+\gamma)
	\left[\left(\sig^{\rm (G)}\right)^{1+\gamma}-E^{\rm (G)}\right]^{-1-\omega}
	-\gamma\frac{p^{\rm (G)}}{\sig^{\rm (G)}}.
	\label{eq:gccg_zeta}
\end{equation}
Substitution into Eq.~\eqref{eq:Vpp_general_eos} gives the second derivative of the potential at the fixed throat:
\begin{equation}
	\begin{aligned}
		\Ueff^{{\rm (G)}\,\prime\prime}(\chi_0)=G_{{\rm v}0}''-8\pi^2\Bigg\{&(\sig_0+2p_0)^2
		+2\sig_0(\sig_0+p_0)\\
		&\times\left[1-2\gamma\frac{p_0}{\sig_0}
		+2\omega(1+\gamma)\left(\sig_0^{1+\gamma}-E^{\rm (G)}_0\right)^{-1-\omega}\right]\Bigg\}.
	\end{aligned}
	\label{eq:gccg_Vpp_static}
\end{equation}
For the void lapse defined in Eq.~\eqref{eq:void_lapse}, together with Eq.~\eqref{eq:void_mass_differential}, the purely geometric term can be written as
\begin{equation}
	G_{{\rm v}0}''=-\frac{4m_{{\rm v}0}}{\chi_0^3}-8\pi\chi_0\rho_{{\rm v}0}'.
	\label{eq:Gvpp0_void}
\end{equation}
Using the numerical prescription established above, Eq.~\eqref{eq:gccg_Vpp_static} is evaluated along the family of static throat radii defined by Eq.~\eqref{eq:horizon_offset_parameterization}. For each admissible value of $\mathcal{E}$, the real branch of $E^{\rm (G)}_0$ is selected from Eq.~\eqref{eq:gccg_static_E_condition}, and the corresponding stability function is then determined. The GCCG-supported thin-shell wormhole is linearly stable whenever
\begin{equation}
	\Ueff^{{\rm (G)}\,\prime\prime}(\chi_0)>0.
	\label{eq:gccg_stability_condition}
\end{equation}

\begin{figure}[htp!]
	\centering
	\includegraphics[width=0.32\linewidth]{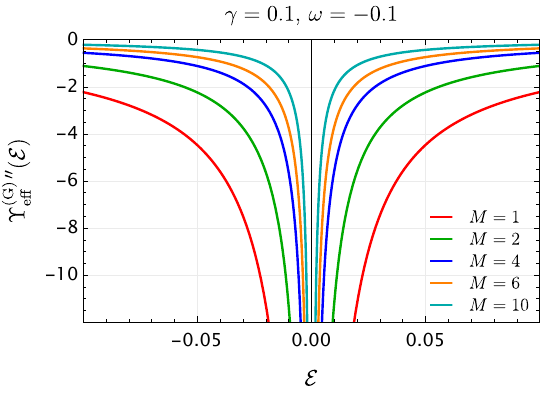}
	\includegraphics[width=0.32\linewidth]{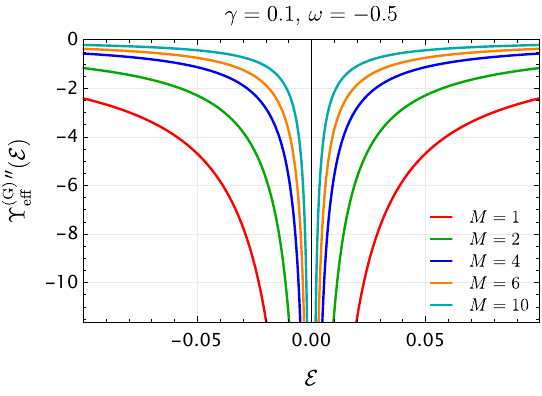}
	\includegraphics[width=0.32\linewidth]{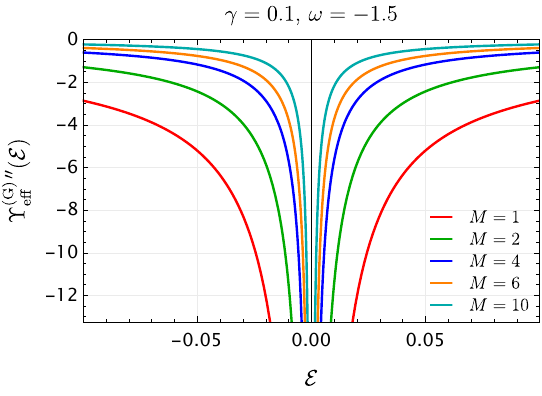}  
	\includegraphics[width=0.32\linewidth]{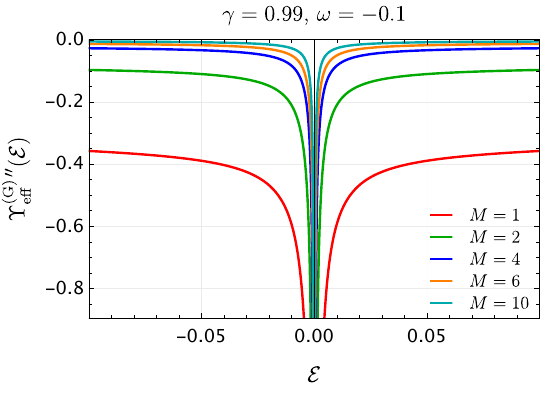}
	\includegraphics[width=0.32\linewidth]{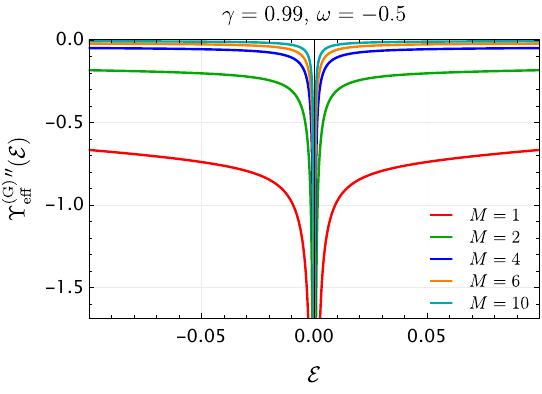}
	\includegraphics[width=0.32\linewidth]{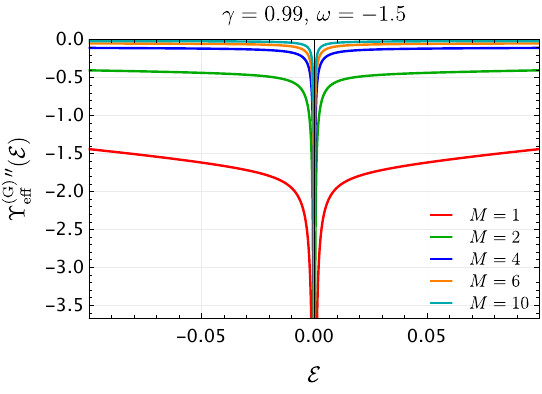}
	\includegraphics[width=0.32\linewidth]{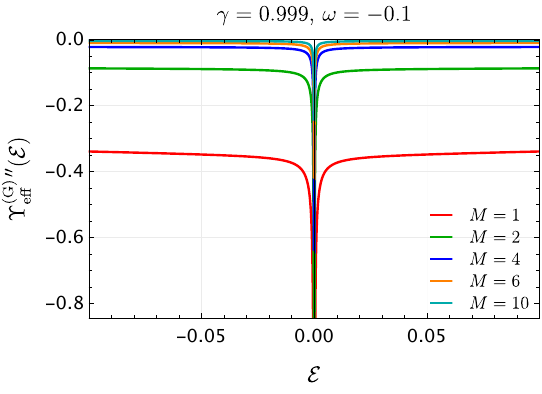}
	\includegraphics[width=0.32\linewidth]{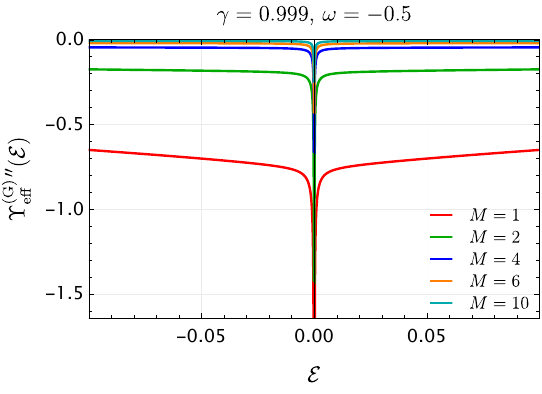}
	\includegraphics[width=0.32\linewidth]{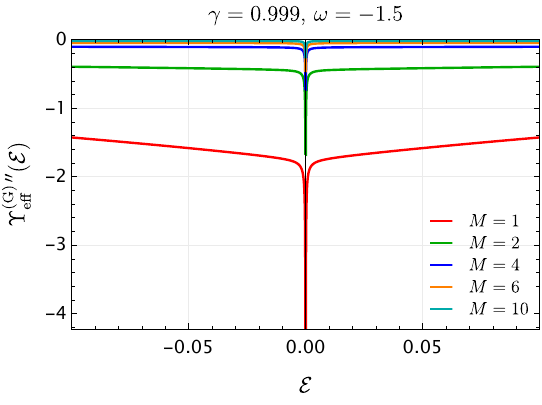}
	\caption{GCCG stability function $\Ueff^{{\rm (G)}\,\prime\prime}$ as a function of the horizon-offset parameter $\mathcal{E}$, with $\chi_0=r_h+|\mathcal{E}|$. In all panels, the density contrast is fixed at $\delta_c=-0.95$, while the curves correspond to $M=1, 2, 4, 6$ and $10$. The panels consider the values $\gamma=0.1$, $0.99$, and $0.999$, and $\omega=-0.1$, $-0.5$, and $-1.5$. For the whole parameter range displayed, the GCCG-supported configurations are linearly unstable.}
	\label{fig:gccg_stability_mass_scan}
\end{figure}

\subsection{Modified cosmic Chaplygin gas (MCCG)}\label{sec:mccg}

For the MCCG ~\cite{SadeghiFarahani2013,SharifMumtaz2014}, the surface equation of state is
\begin{equation}
	p^{\rm (M)}(\chi)=A\sig^{\rm (M)}(\chi)-\frac{1}{\left[\sig^{\rm (M)}(\chi)\right]^\gamma}
	\left[E^{\rm (M)}+
	\left(\left[\sig^{\rm (M)}(\chi)\right]^{1+\gamma}-E^{\rm (M)}\right)^{-\omega}\right],
	\label{eq:mccg_eos}
\end{equation}
where
\begin{equation}
	E^{\rm (M)}=\frac{B^{\rm (M)}}{1+\omega}-1,
	\qquad
	-C<\omega<0.
	\label{eq:mccg_E_definition}
\end{equation}
The radial conservation equation is
\begin{equation}
	\frac{\dd\sig^{\rm (M)}}{\dd\chi}
	=-\frac{2}{\chi}\left\{(1+A)\sig^{\rm (M)}
	-\frac{1}{\left(\sig^{\rm (M)}\right)^\gamma}
	\left[E^{\rm (M)}+
	\left(\left(\sig^{\rm (M)}\right)^{1+\gamma}-E^{\rm (M)}\right)^{-\omega}\right]\right\}.
	\label{eq:mccg_conservation}
\end{equation}
The MCCG effective potential is
\begin{equation}
	\Ueff^{\rm (M)}(\chi)=\Gv(\chi)-4\pi^2\chi^2\left[\sig^{\rm (M)}(\chi)\right]^2.
	\label{eq:mccg_potential}
\end{equation}

At the static throat, $B^{\rm (M)}$ is fixed by
\begin{equation}
	p_0-A\sig_0+\frac{1}{\sig_0^\gamma}
	\left[E^{\rm (M)}_0+
	\left(\sig_0^{1+\gamma}-E^{\rm (M)}_0\right)^{-\omega}\right]=0,
	\label{eq:mccg_static_E_condition}
\end{equation}
with
\begin{equation}
	B^{\rm (M)}_0=(1+\omega)\left(E^{\rm (M)}_0+1\right).
	\label{eq:mccg_B_static}
\end{equation}
Thus, for fixed $A$, $\gamma$, $\omega$, $\chi_0$, and void parameters, the parameter $B^{\rm (M)}$ is determined by the static junction pressure. As with the GCCG case, the MCCG stability is therefore governed by the void geometry and the remaining EoS parameters $(A,\gamma,\omega)$.

The derivative of the MCCG equation of state with respect to the surface density is
\begin{equation}
	\zeta^{\rm (M)}(\chi)=A(1+\gamma)-\gamma\frac{p^{\rm (M)}}{\sig^{\rm (M)}}
	+\omega(1+\gamma)
	\left[\left(\sig^{\rm (M)}\right)^{1+\gamma}-E^{\rm (M)}\right]^{-1-\omega}.
	\label{eq:mccg_zeta}
\end{equation}
Therefore, the second derivative of the effective potential at the fixed throat is
\begin{equation}
	\begin{aligned}
		\Ueff^{{\rm (M)}\,\prime\prime}(\chi_0)=G_{{\rm v}0}''-8\pi^2\Bigg\{&(\sig_0+2p_0)^2
		+2\sig_0(\sig_0+p_0)\\
		&\times\left[1+2A(1+\gamma)-2\gamma\frac{p_0}{\sig_0}
		+2\omega(1+\gamma)\left(\sig_0^{1+\gamma}-E^{\rm (M)}_0\right)^{-1-\omega}\right]\Bigg\}.
	\end{aligned}
	\label{eq:mccg_Vpp_static}
\end{equation}
Note that the term $1+2\zeta^{\rm (M)}(\chi_0)$ in the square bracket has been expanded explicitly. The MCCG-supported static configuration is linearly stable whenever
\begin{equation}
	\Ueff^{{\rm (M)}\,\prime\prime}(\chi_0)>0.
	\label{eq:mccg_stability_condition}
\end{equation}

Equations~\eqref{eq:gccg_Vpp_static} and \eqref{eq:mccg_Vpp_static} are the main local stability criteria of the paper. They show explicitly how the void geometry enters through $m_{{\rm v}0}$, $\rho_{{\rm v}0}$, and $\rho_{{\rm v}0}'$, while the Chaplygin sector enters through the static values $E_0^{\rm (G)}$ and $E_0^{\rm (M)}$, or equivalently through the fixed parameters $B_0^{\rm (G)}$ and $B_0^{\rm (M)}$.

\begin{figure}[htp!]
	\centering
	\includegraphics[width=0.32\linewidth]{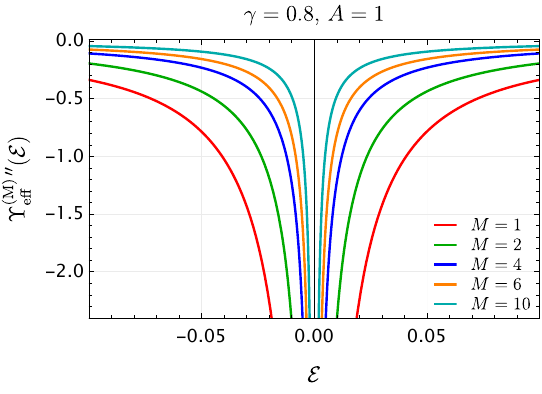}
	\includegraphics[width=0.32\linewidth]{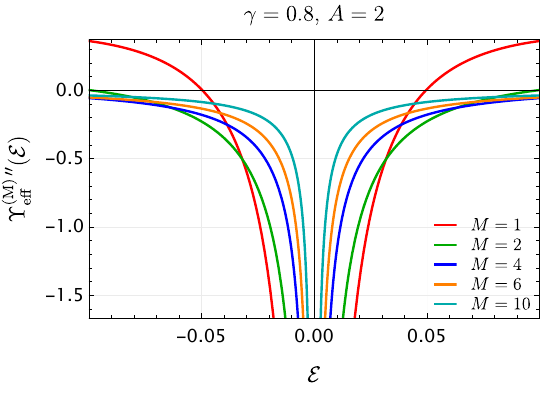}
	\includegraphics[width=0.32\linewidth]{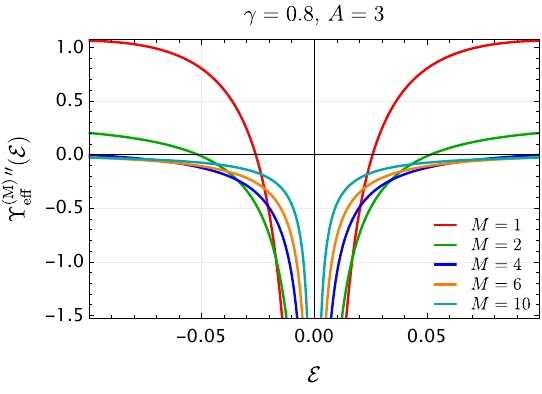}    
	\includegraphics[width=0.32\linewidth]{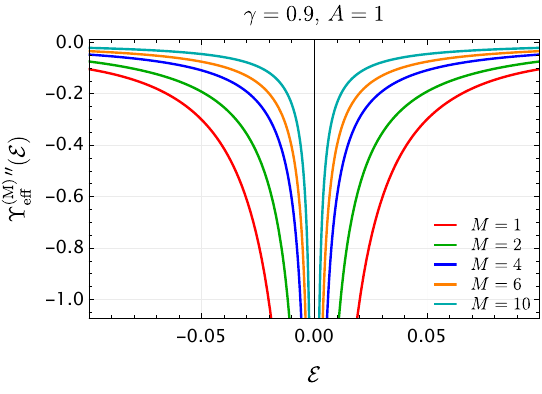}
	\includegraphics[width=0.32\linewidth]{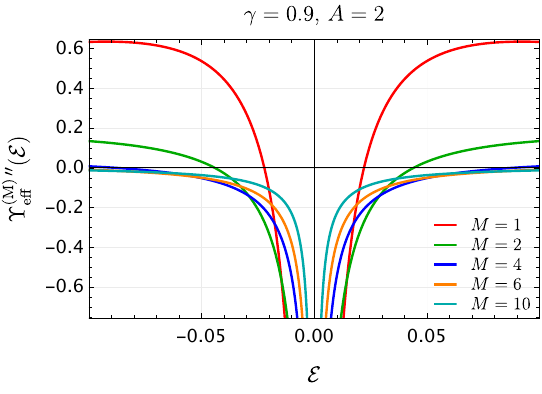}
	\includegraphics[width=0.32\linewidth]{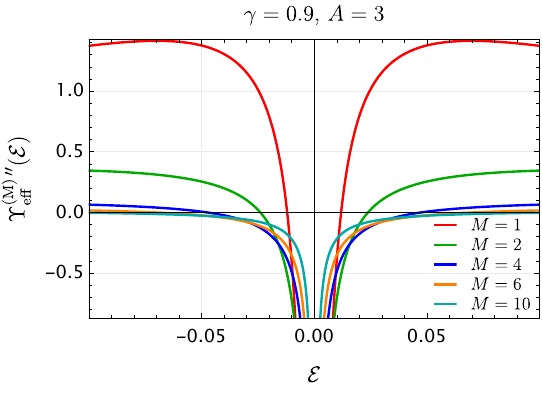}
	\includegraphics[width=0.32\linewidth]{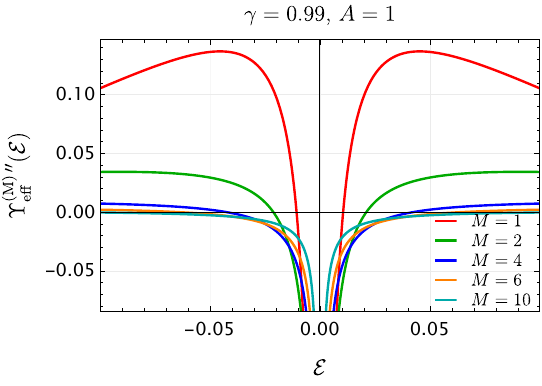}
	\includegraphics[width=0.32\linewidth]{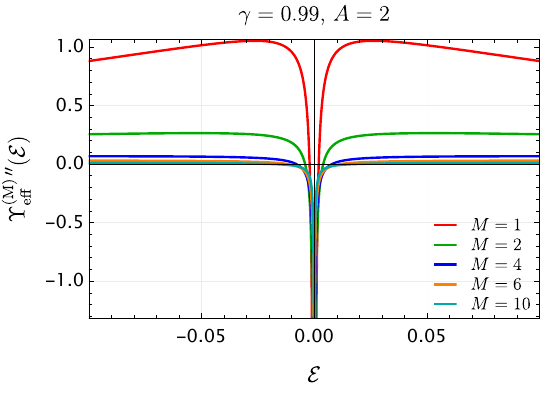}
	\includegraphics[width=0.32\linewidth]{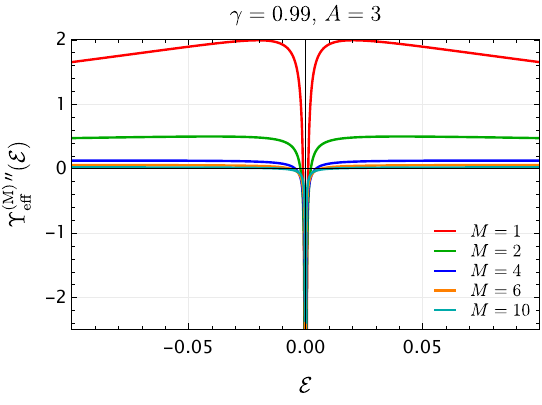}
	\caption{MCCG stability function $\Ueff^{{\rm (M)}\,\prime\prime}$ as a function of the horizon-offset parameter $\mathcal{E}$, with $\chi_0=r_h+|\mathcal{E}|$. In all panels, the density contrast and the cosmic Chaplygin parameter are fixed at $\delta_c=-0.95$ and $\omega=-0.5$, respectively, while the curves correspond to $M=1, 2, 4, 6$ and $10$. The panels consider the values $\gamma=0.8$, $0.9$, and $0.99$, and $A=1$, $2$, and $3$. The horizontal zero level separates stable configurations, $\Ueff^{{\rm (M)}\,\prime\prime}>0$, from unstable ones, $\Ueff^{{\rm (M)}\,\prime\prime}<0$.}
	\label{fig:mccg_stability_mass_scan}
\end{figure}

\section{Conclusions}\label{sec:conclusion}

In this work we have constructed a symmetric thin-shell wormhole inside a cosmic void, using the black-hole-in-void spacetime recently introduced in Ref.~\cite{Lustosa2025} as the seed geometry. The construction is performed in the positive-lapse domain \(r_h<r<r_{++}\) between the black-hole horizon and the cosmological-like horizon generated by the de Sitter-like behaviour of the void density profile. The induced geometry on the shell, the extrinsic curvature, and the surface stress-energy tensor have been obtained through the standard Darmois--Israel junction formalism.

The surface energy density is necessarily negative for the symmetric cut-and-paste configuration, a direct consequence of the flare-out condition. Hence the weak and dominant energy conditions are always violated on the shell, independently of the equation of state chosen for the exotic matter. The null and strong energy-condition combinations, however, are sensitive to the position of the throat inside the static patch: they are satisfied only in the inner region and are violated as the throat approaches the cosmological-like horizon, where the shell becomes tension-dominated.

	The thermodynamic analysis reveals that the static shell possesses a well-defined local Unruh temperature \(T_{\rm shell}=|G_{\rm v}'|/(4\pi\sqrt{G_{\rm v}})\). A first law for the shell fluid is derived, showing that the shell entropy is constant along a sequence of static configurations with fixed ADM mass. When the mass is allowed to vary, the first law yields a differential relation between the shell entropy and the mass, which can be integrated once the equilibrium sequence \(\chi(M)\) is known. Moreover, the shell entropy is directly related to the black-hole horizon entropy through \(\dd S_{\rm shell}= (8\pi/|G_{\rm v}'|) T_{h}\,\dd S_{h}\), reducing to \(\dd S_{\rm shell}\to 2\,\dd S_{h}\) when the throat approaches the black-hole horizon. A similar relation links the shell entropy to the cosmological-like horizon. In the de Sitter limit, the integrated form of the entropy suggests a Smarr-type relation in which the shell contributes additively to the horizon entropy, reinforcing the picture of the wormhole as a composite thermodynamic system whose properties are controlled by the interplay between local exotic matter and the large-scale void environment.

The radial dynamics of the throat has been reduced to a one-dimensional effective potential \(\Ueff(\chi)=\Gv(\chi)-4\pi^{2}\chi^{2}\sigma^{2}(\chi)\). Static configurations correspond to the equilibrium points of this potential, and their linear stability is governed by the sign of \(\Ueff''(\chi_{0})\). For the generalized cosmic Chaplygin gas and the modified cosmic Chaplygin gas equations of state, the Chaplygin parameter \(B\) is fixed by requiring the static pressure to match the junction pressure, leaving the stability test controlled by the remaining parameters \((\gamma,\omega,A)\) and by the void geometry. The numerical analysis shows that GCCG-supported configurations are linearly unstable throughout the explored parameter range, whereas MCCG-supported wormholes can be stable for sufficiently large values of the linear term \(A\), demonstrating how a simple modification of the equation of state can compensate the exotic character of the shell and stabilise the throat.

These results establish a direct link between the large-scale underdense cosmic environment and the local physics of thin-shell wormholes. Several directions for future investigation emerge naturally from this work. First, a systematic scan of the full void parameter space---varying the density contrast \(\delta_{c}\), the void radius \(r_{v}\), and the scale \(r_{s}\)---would determine whether the de Sitter-like environment enlarges or suppresses the stability domains compared with the usual Schwarzschild--de Sitter thin-shell wormholes. Second, the inclusion of back-reaction effects and the study of dynamical evolution beyond linear perturbations would clarify whether the linearly stable configurations identified here remain stable under finite radial oscillations. Third, the near-Nariai limit, where the two horizons merge and the static patch disappears, requires a separate treatment that likely goes beyond the classical thin-shell formalism and may call for quantum-gravitational corrections. Fourth, the connection between the thermodynamic description developed here and the Hawking evaporation of the horizons could provide an effective framework for studying the life-time and observational signatures of void-embedded wormholes. Finally, extending the present construction to rotating or higher-dimensional void geometries would open the possibility of linking these results to recent discussions of wormhole phenomenology in large-scale structure and gravitational-wave astronomy.

\begin{acknowledgments}
EO thanks the Fundação Cearense de Apoio ao Desenvolvimento Científico e Tecnológico (FUNCAP), through grant BP6-0241-00335.01.00/25.
FSNL acknowledges funding from the Fundacão para a Ciência e a Tecnologia (FCT) through the research grant UID/04434/2025 and the FCT Scientific Employment Stimulus contract with reference CEECINST/00032/2018. 
\end{acknowledgments}

\bibliographystyle{apsrev4-2}

\bibliography{ref}

\end{document}